\documentclass{sig-alternate}
\usepackage{mathptmx} 

\usepackage{cite}
\usepackage{amsmath,amssymb,amsfonts}
\usepackage{algorithmic}
\usepackage{graphicx}
\usepackage{textcomp}
\usepackage[table]{xcolor}
\usepackage{fancyhdr}
\usepackage[hyphens]{url}

\usepackage{authblk}

\makeatletter
\renewcommand\AB@affilsepx{, \protect\Affilfont}
\makeatother

\usepackage[final]{microtype}
\usepackage{flushend}
\usepackage{booktabs}
\usepackage{array}
\usepackage[subrefformat=parens,labelformat=parens]{subfig}   
\usepackage{tablefootnote}
\usepackage{tabularx}
\usepackage{diagbox}
\usepackage{multirow}

\newcommand{\todo}[1]{}
\renewcommand{\todo}[1]{{\color{red} TODO: {#1}\\}}
\newcommand{\RED}[1] {#1} 

\newcommand\lword[1]{\leavevmode\nobreak\hskip0pt plus\linewidth\penalty50\hskip0pt plus-\linewidth\nobreak{#1}}

\usepackage{soul}
\usepackage{amsmath}
\usepackage{amssymb}
\usepackage{pifont}
\newcommand{\cmark}{\ding{51}}%
\newcommand{\xmark}{\ding{55}}%
\usepackage{slashbox}
\usepackage{xspace}
\usepackage[bookmarks=true,breaklinks=true,letterpaper=true,colorlinks,linkcolor=black,citecolor=blue,urlcolor=black]{hyperref}
\usepackage{xfrac}
\usepackage{enumitem}

\renewcommand{\paragraph}[1]{\textbf{#1} }
\newcommand{\x}{$\times$}

\newcommand{\YES}{\textcolor{green}{\cmark}}
\newcommand{\NO}{\textcolor{red}{\xmark}}

\newcommand\blfootnote[1]{%
  \begingroup
  \renewcommand\thefootnote{}\footnote{#1}%
  \addtocounter{footnote}{-1}%
  \endgroup
}
\newcommand{\GAP}{} 

\def\BibTeX{{\rm B\kern-.05em{\sc i\kern-.025em b}\kern-.08em
    T\kern-.1667em\lower.7ex\hbox{E}\kern-.125emX}}

\fancypagestyle{firstpage}{
  \fancyhf{}

  \fancyfoot[C]{\thepage}
}

\title{S2TA: Exploiting Structured Sparsity for\\Energy-Efficient Mobile CNN Acceleration
}

\author[*$\dagger$]{Zhi-Gang Liu}   
\author[*$\dagger$]{Paul N. Whatmough}
\author[+]{Yuhao Zhu}
\author[*]{Matthew Mattina} 
\affil[*]{Arm ML Research Lab, Boston, MA, USA} 
\affil[+]{University of Rochester, Rochester, NY, USA} 

\ifx\figurename\undefined \def\figurename{Figure}\fi
\renewcommand{\figurename}{Fig.}
\renewcommand{\paragraph}[1]{\textbf{#1} }

\newcommand{\Sect}[1]{Sec.~\ref{#1}}
\newcommand{\Fig}[1]{Fig.~\ref{#1}}
\newcommand{\Tbl}[1]{Tbl.~\ref{#1}}

\newcommand{\mode}[1]{\underline{\textsc{#1}}\xspace}

\newcommand{\no}[1]{#1}
\renewcommand{\no}[1]{}
\newcommand{\RNum}[1]{\uppercase\expandafter{\romannumeral #1\relax}}

\begin{document}
\maketitle
\thispagestyle{firstpage}
\pagestyle{plain}

\begin{abstract}

Exploiting sparsity is a key technique in accelerating quantized convolutional neural network (CNN) inference on mobile devices.
Prior sparse CNN accelerators largely exploit unstructured sparsity and achieve significant speedups. Due to the unbounded, largely unpredictable sparsity patterns, however, exploiting unstructured sparsity requires complicated hardware design with significant energy and area overhead, which is particularly detrimental to mobile/IoT inference scenarios where energy and area efficiency are crucial.

We propose to exploit \textit{structured} sparsity, more specifically, Density Bound Block (DBB) sparsity for both weights and activations.
DBB block tensors bound the maximum number of non-zeros per block. DBB thus exposes statically predictable sparsity patterns that enable lean sparsity-exploiting hardware and efficient memory access.
We propose new hardware primitives to implement DBB sparsity for (static) weights and (dynamic) activations, respectively, with very low overheads.

Building on top of the primitives, we describe S2TA, a systolic array-based CNN accelerator that exploits joint weight and activation DBB sparsity and new dimensions of data reuse unavailable on the traditional systolic array.
S2TA in 16nm achieves more than 2$\times$ speedup and energy reduction compared to a strong baseline of a systolic array with zero-value clock gating, over five popular CNN benchmarks.
Compared to two recent non-systolic sparse accelerators, Eyeriss v2 (65nm) and SparTen (45nm), S2TA in 65nm uses about 2.2$\times$ and 3.1$\times$ less energy per inference, respectively.

\end{abstract}



\section{Introduction}
\label{sec:intro}
\GAP

\blfootnote{$^\dagger$ Denotes equal contribution.}



Convolutional neural network (CNN) inference has quickly become an important workload in (ultra) low-power mobile ~\cite{euphrates,asv} and 
IoT/embedded~\cite{kodali-iccd17,fedorov2019sparse,tinylstm,micronets-mlsys2021} devices. CNN accelerators are now a standard component in mobile SoCs~\cite{smiv,smiv-jssc21,hansen-icpr20,zhu2018mlsoc}, where
8-bit integer (INT8) CNN inference is the most widely used~\cite{8bit-warden} due to the stringent requirements on energy efficiency (TOPS/W) and area efficiency (TOPS/mm$^2$).

A common strategy to improve CNN accelerator efficiency is to exploit sparsity, as zeros in the data tensors (both weights and activations) reduce the theoretical compute and storage requirement significantly. 
Zeros in CNNs are statistically distributed in a random pattern.
Exploiting the random sparsity, which is also commonly referred to as \textit{unstructured} sparsity, has been the main focus of sparse hardware accelerators to date~\cite{eie,scnn,laconic_isca19,sparten-micro19,eyeriss_v2}. Exploiting unstructured sparsity, however, requires complex hardware structures that introduce significant area and energy overhead, which are particularly detrimental to mobile/embedded inference scenarios. 


Exploiting unstructured sparsity introduces hardware overheads due to additional buffers used in data manipulation.
There are two fundamental approaches to supporting random sparsity.
The first is the inner-product style~\cite{smt-sa,eyeriss_v2} which requires an \textit{operand gather} stage to evenly distribute the unpredictable workload to the PEs to perform the MAC operations.
The second, is the outer-product style which has more conventional operand distribution and MAC operations, but then requires a \textit{result scatter} stage to spread the non-contiguous results of the individual MACs over the output feature maps. 
Both of these approaches introduce significant hardware overheads in the form of large buffers that drastically degrade the energy and area-efficiency of the accelerator.

We quantitatively show that the additional buffers required to exploit unstructured sparsity,
which may seem like nitty-gritty engineering details, in fact increase the energy per MAC of a baseline (systolic array) dense accelerator by 71\%. As a result, random sparse accelerators published to date have demonstrated high speedup, but due to hardware overheads, have limited energy and area efficiency gains.

We argue that the sparsity-exploiting microarchitecture structures must be lightweight for it to be beneficial to exploit sparsity at all in mobile/embedded CNN accelerators. Otherwise, the energy/area overheads can easily eclipse the speedup gains.  To that end, we propose to exploit \textit{structured sparsity}, which has regular sparsity patterns that allow the hardware additions required to be very lean. In particular, we focus on Density Bound Block (DBB)~\cite{sta-cal2020}
format, which
tiles data tensors into blocks, and then introduces a bound on the maximum number of non-zero elements per block.
DBB overcomes both of the challenges with random sparsity.
Firstly, the maximum number of MAC operations is fixed, which significantly achieves high PE utilization \textit{without} operand buffers.
Secondly, the blocked data limits the location of output results, eliding distributed accumulators.

We first propose two 
primitives: W-DBB and A-DBB, which exploit DBB in weight 
and activation sparsity, respectively. Weight sparsity is statically known, and thus the hardware design is relatively straightforward. Activation sparsity, however, is dynamic. We propose a dynamic pruning scheme co-designed with a novel time-unrolled architecture to implement A-DBB.

Building on 
W-DBB and A-DBB, we describe how to integrate the two building blocks into a fully-sparse CNN accelerator that efficiently exploits sparsity in both weights and activations. In particular, we focus on the well-known systolic array architecture, popular in mobile/embedded inference scenarios due to its extreme efficiency arising from high data reuse. We show that both W-DBB and A-DBB can be easily integrated into a classic SA architecture by introducing a tensor PE (TPE) design
enabling further efficiency gains. 
\begin{figure}[t]
    \centering
    \includegraphics[width=0.9\columnwidth]{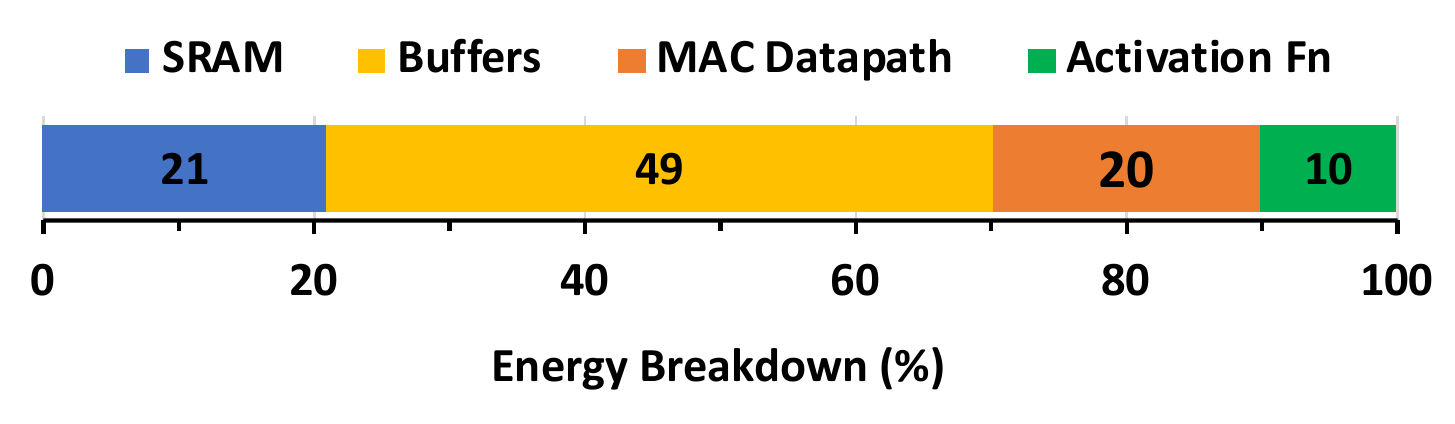}
    \caption{
    Energy breakdown for a conventional dense INT8 systolic array accelerator, with typical 50\% CNN sparsity.
    The INT8 MAC datapath itself is very compact (consuming only 20\% of the energy), while the operand/result buffers dominate.
    }
    \label{fig:sa-energy-pie}
\end{figure}

In summary, this paper makes the following contributions:
\begin{itemize}
\item \textbf{Quantifying Overhead of Exploiting Unstructured Sparsity} We provide the first quantitative analysis of the hardware overheads (area and energy)
of exploiting unstructured sparsity. We show that the additional operand and accumulator buffers
introduce about {50\% and 10\%} energy and area overhead compared to the datapath of a baseline dense INT8 CNN accelerator.
These overheads are amortized in floating-point designs ~\cite{tensordash,SIGMA,smt-sa}, where the datapath power is much higher, but in the mobile INT8 case, these overheads eclipse the gains.

\item \textbf{Joint Weight/Activation DBB Sparsity}
We propose an architecture that exploits DBB sparsity in both weights and activations.
This is non-trival, because while weight sparsity is known offline, activation sparsity is not known until runtime and can vary wildly.
We propose Dynamic Activation Pruning (DAP), which co-designs training time activation DBB pruning and novel run-time hardware support.
DAP compresses activations by 2--3$\times$ with negligible impact on test accuracy.

\RED{\item \textbf{Time-Unrolled Variable DBB Sparsity}
We propose a new time-unrolled microarchitecture to serialize the processing of MACs in a DBB block across time. }   

\item \textbf{Structured Sparse Tensor Accelerator (S2TA)}
We show that joint DBB weight and activation sparsity can be efficiently incorporated into CNN accelerators. As a case study, we target the systolic array template, by extending the traditional scalar PE into a tensor PE (TPE) that consumes compressed DBB data blocks.




\item \textbf{Evaluation in 16nm and 65nm}
We implement the S2TA design
in both 16nm and 65nm technology.
On a range of popular models (AlexNet, MobileNetv1, VGG-16, ResNet50v1), S2TA demonstrates 2.08$\times$ lower energy compared to the baseline without DBB.
Compared to the state-of-the-art unstructured sparse accelerator SparTen~\cite{sparten-micro19} and Eyeriss-v2~\cite{eyeriss_v2}, S2TA 
has \RED{2.2$\times$ and 3.1$\times$} lower energy on AlexNet in 65nm technolgy.

\end{itemize}

\section{Motivation} \label{sec:background}

In this section, we review the two fundamental approaches to supporting random sparsity in embedded CNN accelerators: (1) inner-product style with operand gather, and (2) outer-product style with result scatter. 
Both introduce significant hardware overheads that drastically degrade the energy and area-efficiency of the accelerator.

%

\subsection{Sparsity-Exploiting Structures Must be Efficient for Mobile DNN Inference}
\label{sec:sparse_dnn}
\GAP

The simplest way to exploit sparsity in hardware is Zero Value Clock Gating (ZVCG). \Fig{fig:datapaths}b shows how ZVCG simply detects zero operands (weights and activations) and clock-gates the operand and/or result registers to reduce power dissipation. While ZVCG gives a significant reduction in datapath power~\cite{eyerissIsca,minerva}, it does not increase throughput, nor does it reduce the SRAM bandwidth (as the zeros are still stored and read in sequence). More importantly, ZVCG reduces the hardware utilization and, thus, does not improve area efficiency (TOPS/mm$^2$), which is critical for mobile.

Sparse GEMM potentially achieves much higher gains than ZVCG by reducing both the number of MACs that need to be executed and also the memory bandwidth.
To that end, only the non-zero values and a corresponding positional index are stored.
The index encodes the position of each non-zero element in the expanded matrix, via either compressed sparse row/column (CSR/CSC)~\cite{eie,scnn} format or a simple bitmask~\cite{sparten-micro19}.
At runtime, both the SRAM bandwidth, and the number of MACs executed are significantly reduced.

However, implementing fully sparse GEMM on energy and area-constrained INT8 mobile/IoT accelerators is challenging. Basically, removing MACs with zero operands breaks the regular compute pattern and requires complex on-chip buffering and data re-ordering to maximize the hardware utilization. The additional buffers significantly increase the energy and area overhead. 
\Fig{fig:sa-energy-pie} shows the energy breakdown of an INT8 dense systolic array accelerator for a typical CNN layer. The data is obtained from the extracted post-layout power estimation in a 16nm technology node with fully annotated switching activity.

\paragraph{Key Insight} 
The energy consumption of the actual INT8 MAC computation in \Fig{fig:sa-energy-pie} is significantly overshadowed by the buffers used for operands and accumulators. 
Therefore, any sparsity-exploiting scheme must not introduce significant overheads that exacerbate the data buffering cost, as that is already the dominant energy consumer.

Today's sparsity-exploiting mechanisms, unfortunately, introduce significant area and energy overhead for data buffering due to non-trivial data re-ordering, which can be classified into two fundamental categories, as illustrated in \Fig{fig:random-overheads}. 
The first category (\Fig{fig:random-overheads}a) performs a \textit{gather} operation in the front-end to collect matching pairs of non-zero operands before buffering and finally performing MAC computations in the conventional manner. This approach is seen in previous work, including EIE~\cite{eie}, Eyerissv2~\cite{eyeriss_v2}, and SMT-SA~\cite{smt-sa}.
The second approach (\Fig{fig:random-overheads}b) avoids re-ordering operands in the front-end, by computing an outer-product (i.e. multiplying all non-zero weights with all non-zero activations). This approach is seen in accelerators such as SCNN~\cite{scnn} and SparTen~\cite{sparten-micro19}.
However, this requires a \textit{scatter} operation on the partial products using a very large number of read-modify-write accumulators to store the output activations.
The rest of this section quantifies the overhead associated with these two approaches.



\subsection{Overhead 1: Operand Gather Structure}
\label{sec:background:oh1}
\GAP


\begin{figure}[t]
\centering
\hspace*{-0.2cm}
\includegraphics[width=0.96\columnwidth]{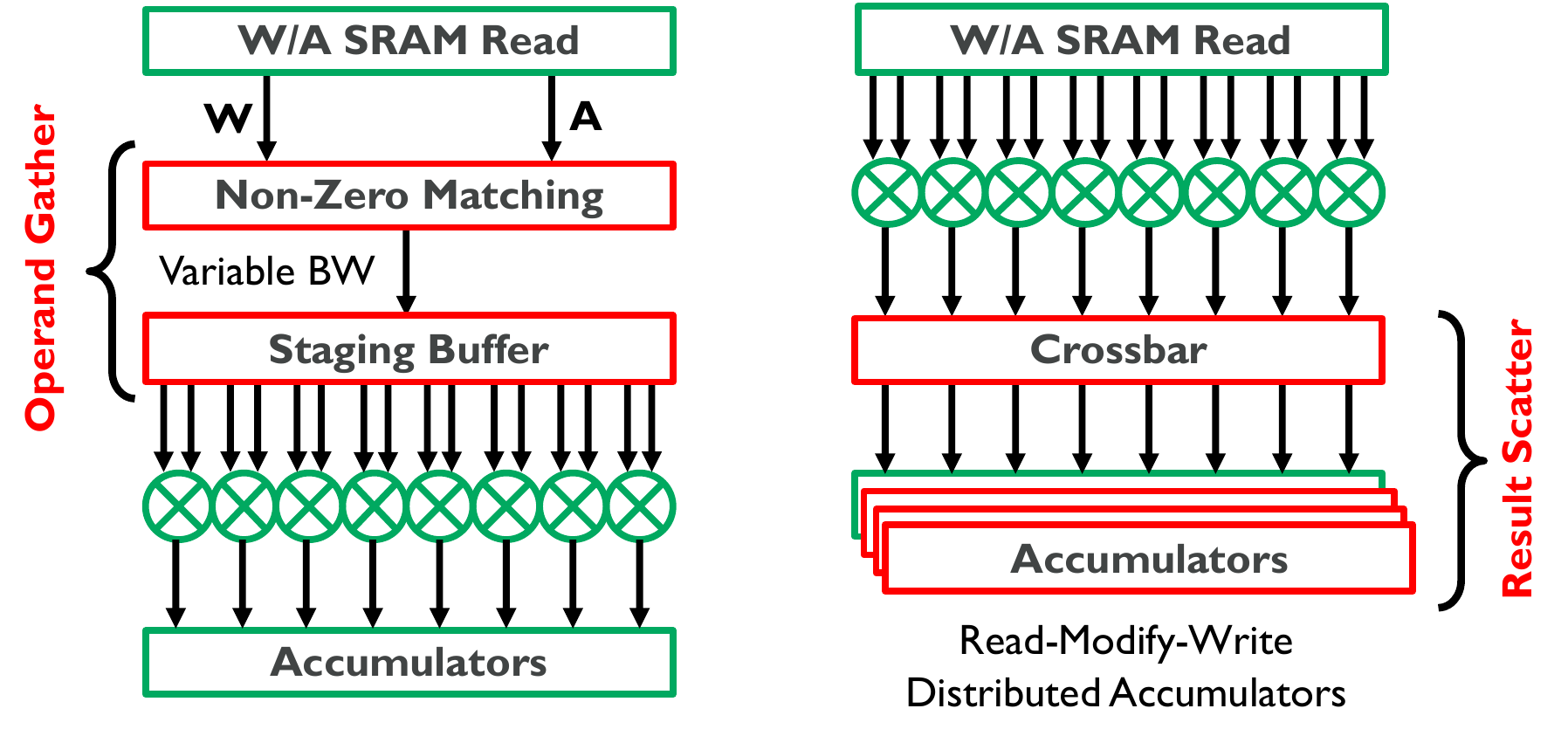}
(a) Inner-Product (Eyerissv2)
\hspace*{0.05cm}
(b) Outer-Product (SCNN)
\vspace{4pt}
\caption{
Hardware structures for sparse GEMM require explicit data re-ordering, which introduces overheads (blocks in red), in the form of either (a) an operand gather stage before the MAC compute (SMT-SA~\cite{smt-sa}), or (b) a result scatter with a distributed accumulator (SCNN~\cite{scnn}).
}
\label{fig:random-overheads}
\vspace{8pt}
\end{figure}

\begin{figure}[t]
    \centering
    \hspace*{-0.3cm}
    \includegraphics[width=1.0
    \columnwidth]{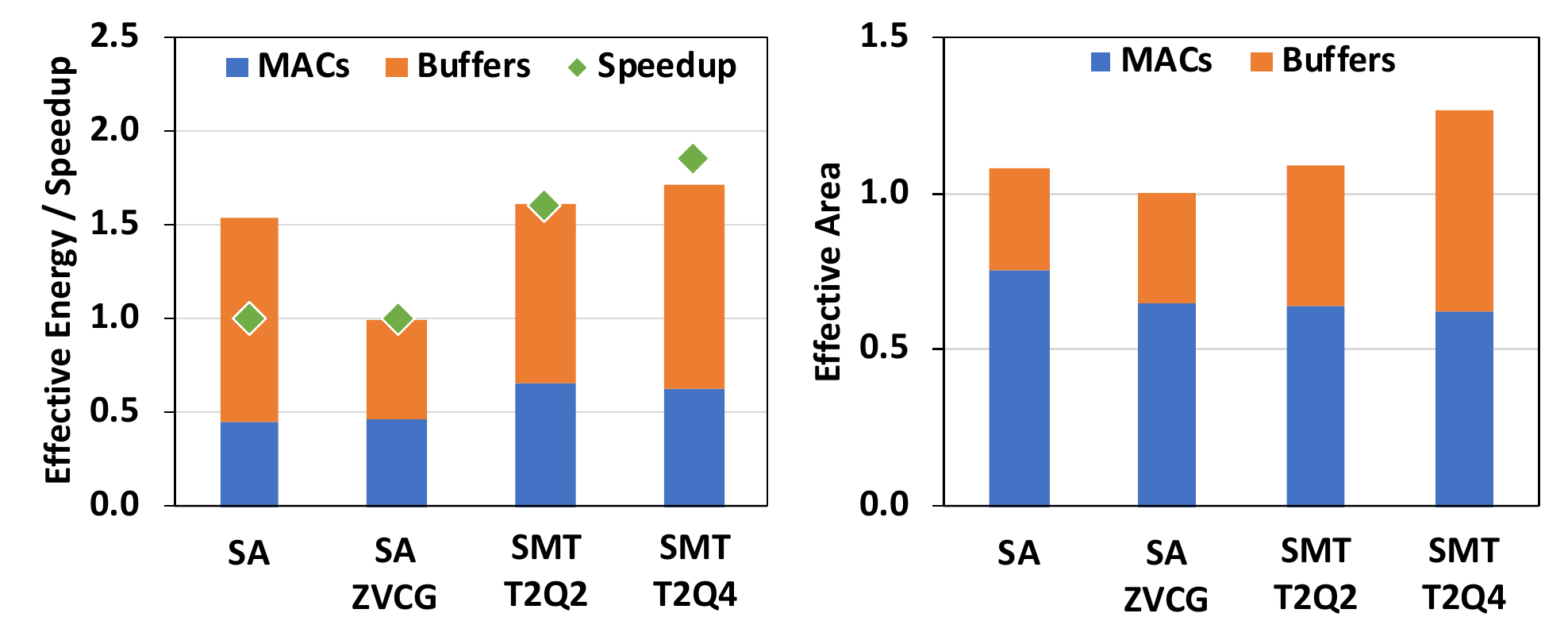}
    \vspace{3pt}
    \caption{
    Effective energy/area and speedup, with breakdown by component for INT8 systolic array variants processing a typical convolution with 50\% weight and activation sparsity.
    The SMT, our INT8 re-implementation of~\cite{smt-sa}, achieves speedup, but buffering overheads actually result in worse energy efficiency than even the dense case.
    }
    \label{fig:array-energy-area}
\end{figure}

The weight ($W$) and activation ($A$) tensors have independent random sparsity patterns.
Thus, during execution, we must walk the indexes to find matching pairs of non-zero positions to pass to the MAC units.
The number of matches varies wildly depending on the position and the input data, which gives rise to an unpredictable number of MACs in any single cycle, leading to variable, unbalanced PE utilization at run time.



To keep the MAC utilization high, 
this load imbalance is usually evened out using a data staging buffer, which collects the matched operand pairs and packs them into groups of a fixed size that matches the datapath width~\cite{tensordash,SIGMA}.
The hardware supporting this approach is shown in \Fig{fig:random-overheads}a. While the non-zero operand matching itself may be of a reasonable cost~\cite{sparten-micro19}, the data staging buffer (typically FIFO) introduces high energy and area overheads, which are especially significant for low energy/area INT8 mobile/embedded accelerators.

\Fig{fig:array-energy-area} quantifies the energy and area overhead of the FIFOs used for distributing matching pairs for INT8 operands. We compare the energy and area across four designs: a dense systolic array (\mode{SA}), a systolic array with ZVCG optimization (\mode{SA-ZVCG}), and two variants of \mode{SA-SMT}, a recent systolic array that exploits random sparsity using operand FIFOs~\cite{smt-sa}. The two \mode{SA-SMT} variants differ by their FIFO depths; one uses 2-entry FIFOs (\mode{SMT-T2Q2}) and the other uses 4-entry FIFOs (\mode{SMT-T2Q4}).
The PPA is obtained for a typical convolution layer with 50\% weight and activation sparsity. 
The energy and area are broken down into the two key SA components: MACs (compute) and the on-chip buffers. 

\paragraph{Key Insight}
We find that the two SA-SMT variants achieve 1.6$\times$ and 1.8$\times$ speedup.
However, this requires additional buffering for load balancing, introducing significant area and energy overheads. 
Overall, despite the speedup achieved, \mode{SMT} shows nearly 50\% \textit{higher} power and roughly same area to \mode{SA-ZVCG}, with INT8 operands required for mobile. 




\begin{table}[t]
\caption{
Comparison of PE buffer sizes per INT8 MAC.
Outer-product style accelerators (SCNN and SparTen) require large buffers per MAC compared to a baseline systolic array.
}
\vspace{2pt}
\centering
\scriptsize
\begin{tabular}{l l l l}
\toprule
\textbf{Architecture}                & \textbf{Operands}   & \textbf{Accumulators}  & \textbf{Total}    \\
\midrule
SCNN~\cite{scnn}                     & 1.28 KB     & 0.375 KB             & 1.65 KB         \\ 
SparTen~\cite{sparten-micro19}       & 864 B       & 128 B              & 0.99 KB           \\ 
Eyeriss v2~\cite{eyeriss_v2}         & 165 B       & 40 B                & 205 B            \\ 
SA-SMT$^1$~\cite{smt-sa}             & 16 B         & 4 B          & 20 B            \\ 
Systolic Array$^1$~\cite{tpu-isca}       & 2 B         & 4 B          & 6 B             \\ 
\RED{S2TA-W$^2$ (Ours)}                        & \RED{0.375 B}      & \RED{0.5 B}         & \RED{0.875 B}          \\
\textbf{S2TA-AW$^3$} (Ours)                       & 0.75 B      & 4 B         & 4.75 B          \\
\bottomrule
\end{tabular}
\\
\vspace{2pt}
\footnotesize
$^1$Our INT8 implementation. \quad
\RED{$^2$4/8 W-DBB (4\x4\x4\_4\x8)}.\\ 
\RED{$^3$Time-Unrolled AW-DBB (8\x4\x4\_8\x8, BZ=8).}
\label{tab:buffers}
\end{table}

\subsection{Overhead 2: Result Scatter Structure}
\label{sec:background:oh2}
\GAP


A common alternative to the gather structure for matched operands, is to multiply every non-zero weight with every non-zero activation (i.e., an outer-product). In this approach, however, the MAC operations being executed in parallel correspond to different, typically non-contiguous elements in the output feature map. Therefore, the results of individual MACs must be properly distributed to the right elements in the output feature map, using a scatter operation.

Distributing the results is expensive because, instead of using a simple local output stationary accumulator register in the dense accelerator, one must construct a very large accumulator buffer using FF or SRAM, where each accumulator requires a read-modify-write operation to accumulate the partial sum in the correct place. The \RED{area and power} overhead of this large accumulator buffer is, again, significant for INT8 inference accelerators, where the datapath logic is relatively cheap compared to the buffers (\Fig{fig:sa-energy-pie}).

\Tbl{tab:buffers} quantifies the overhead of the additional buffers that implement the result scatter function. Specifically, the table compares the buffer size per PE of the baseline systolic array and a range of different accelerators.
Of these accelerators, SCNN~\cite{scnn} and SparTen~\cite{sparten-micro19} implement a scatter structure. While the systolic array requires only 2B for storing operands and 4B for storing the accumulators per PE, the storage requirement is significantly higher for other accelerators: e.g., SCNN~\cite{scnn} reports a total buffer size of 1.65 KB per PE, and SparTen~\cite{sparten-micro19} requires about 1 KB per PE. 

\section{Main Idea and Design Overview}
\label{sec:idea}
\GAP

\begin{figure}[t]
\centering
\hspace*{-0.25cm}
\includegraphics[width=0.4\textwidth]{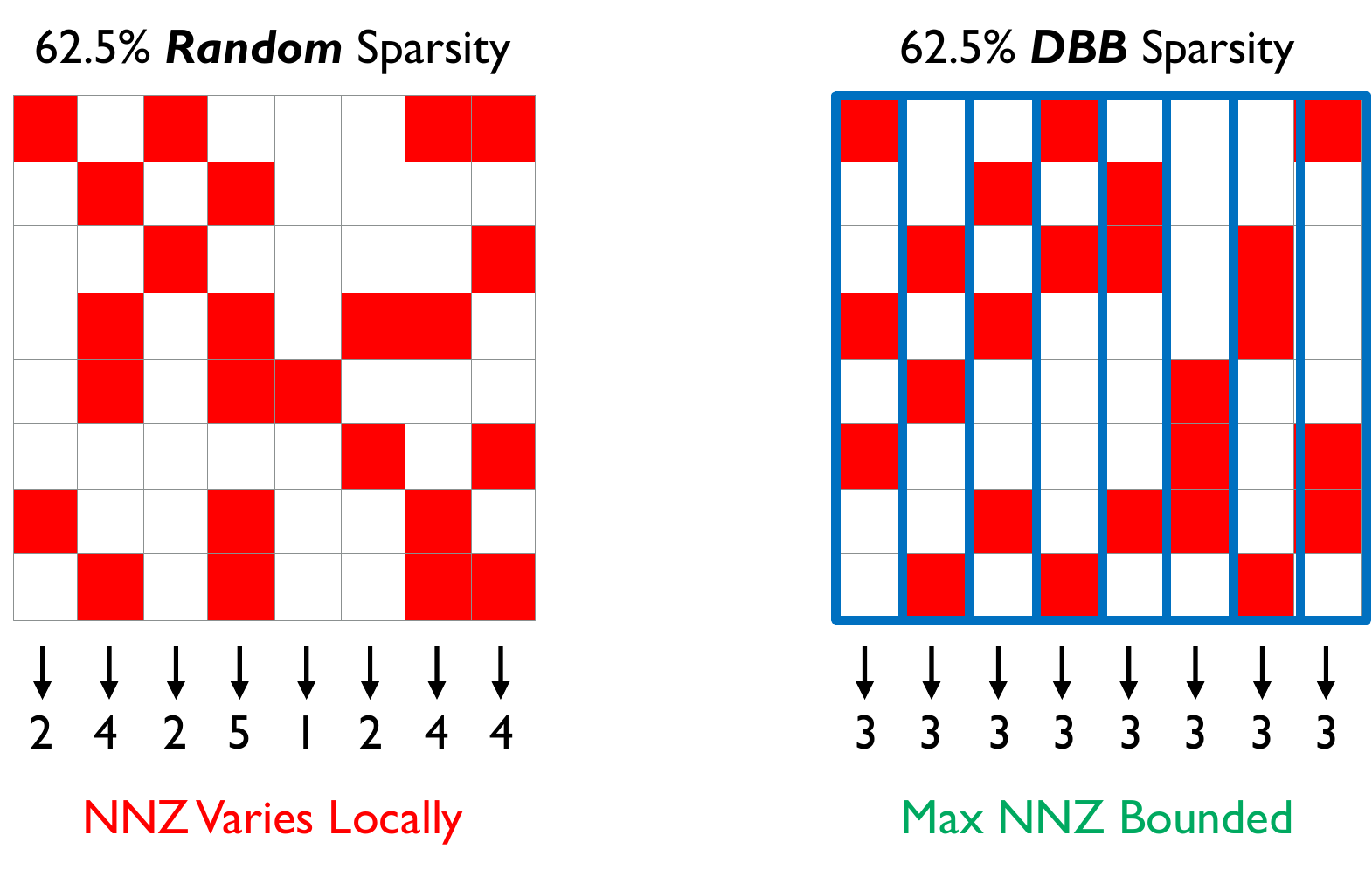}
\\
\footnotesize
\hspace{0pt}
(a) Unstructured Sparse Tensor
\hspace{20pt}
(b) DBB Sparse Tensor\hspace{8pt}
\hspace{2pt}
\vspace{4pt}
\normalsize
\caption{
Comparison of (a) an unstructured sparse tensor, and (b) a density bound block (DBB) sparse tensor. DBB constrains the maximum number of non-zeros (NNZ) per block, such that the max workload is known at design time. The NNZ shown is 3, and the block size (BZ) is 8.
}
\label{fig:sparse_mats}
\end{figure}

This section first introduces the DBB structured sparsity, which we argue enables an efficient sparsity-exploiting architecture (\Sect{sec:idea:dbb}). We then highlight the rationale of our architecture support for DBB, with the detailed designs to follow in the subsequent sections (\Sect{sec:idea:ov}).


\subsection{Exploiting Structured Sparsity}
\label{sec:idea:dbb}
\GAP

The sparsity-exploiting structures must be lightweight to be at all beneficial in exploiting sparsity in mobile CNN accelerators. To that end, we propose to exploit \textit{structured sparsity}, which has regular sparsity patterns that allow the hardware design to be lean. In particular, we focus on leveraging the Density Bound Block (DBB) sparsity. 
DBB essentially divides an (activation/weight) tensor into blocks and sets the upper bound of the number of non-zero (NNZ) elements in each block. \Fig{fig:sparse_mats} compares unstructured sparsity with DBB sparsity. While the overall sparsity level is the same, DBB sparsity constrains the maximum number of NNZs in a block such that the maximum workload is known at design time.


DBB provides two benefits. First, processing DBB blocks in order ameliorates the load imbalance problem and removes the distributed accumulator problem encountered with unstructured sparsity, avoiding the energy- and area-hungry buffers.
Second, DBB lends itself to a simple hardware design based around the movement of small DBB data blocks, as we will illustrate with our S2TA accelerator (\Sect{sec:npu}).
\RED{This compact and efficient hardware architecture is possible because results are naturally generated in sequential blocks; no data reordering is required.}


\begin{figure}[t]
\centering
\includegraphics[width=0.49\textwidth]{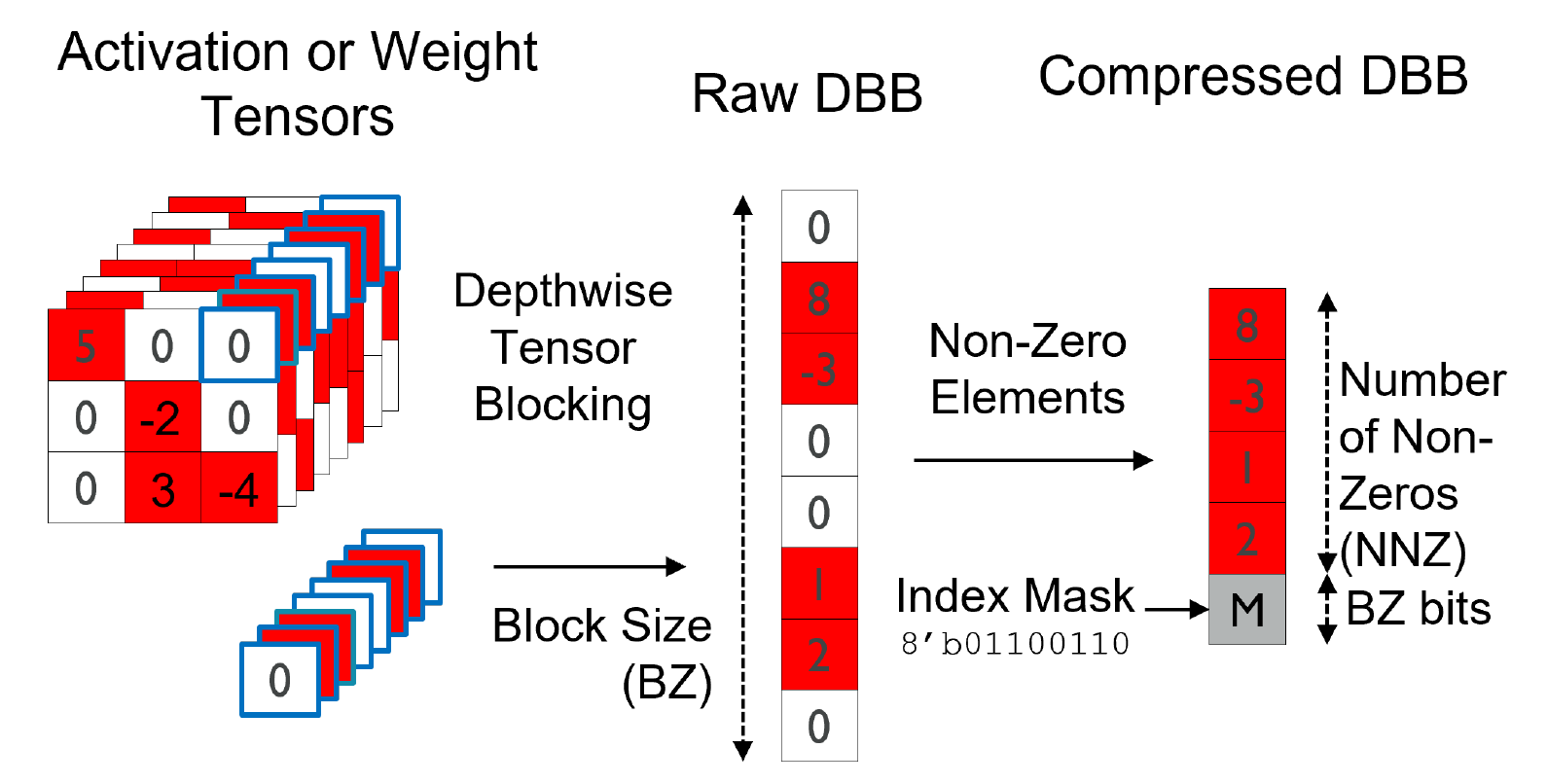}
\vspace{-3pt}
\caption{
A $4/8$ DBB example, where BZ=8 and NNZ=4. The tensor is blocked along the channel dimension.
The block is compressed by storing only the non-zero elements and their positional index bitmask (M). \RED{The same compression applies to weights and activation tensors.}
}
\label{fig:dbb_compress}
\end{figure}

\begin{figure*}[t]
\centering
\includegraphics[width=2.08\columnwidth]{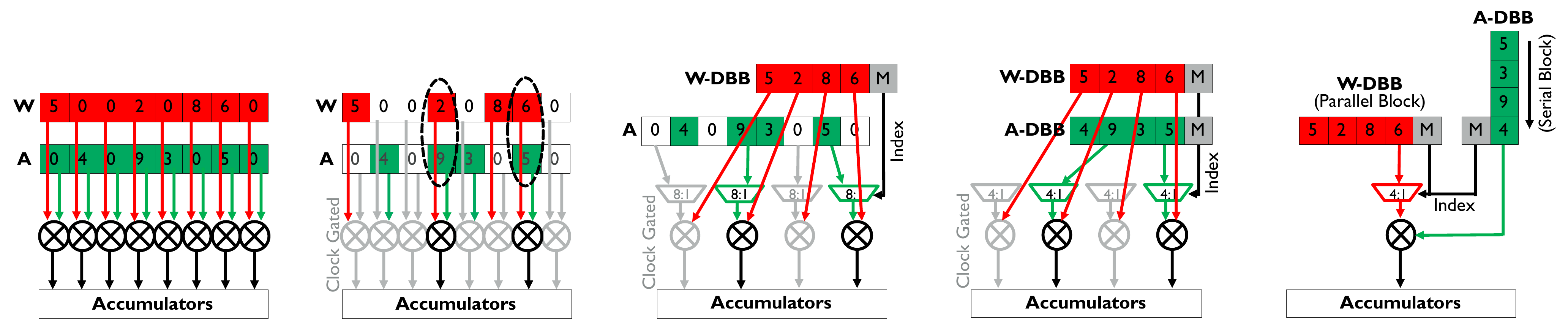}
\\
\vspace{2pt}
\footnotesize
\hspace*{0.5cm}
(a) Dense (DP8)
\hspace*{1.0cm}
(b) ZVCG (DP8)
\hspace*{1.2cm}
(c) W-DBB (DP4M8)
\hspace*{0.7cm}
(d) A/W-DBB (DP4M4)
\hspace*{0.4cm}
(e) A/W-DBB (DP1M4)
\hspace*{0.0cm}
\vspace{8pt}
\normalsize 
\caption{
Parallel sparse MAC hardware. 
(a) Dense 8-MAC vector dot product (DP8). 
(b) Dense DP8 with zero-value clock gating (ZVCG) reduces power by exploiting random sparsity.
(c) $4/8$ weight DBB (W-DBB) sparsity uses only 4 MACs (DP4), with an 8:1 mux (M8) for each (DP4M8).
(d) A/W-DBB uses a 4-way mux (DP4M4), but is limited to fixed DBB sparsity ratios.
(e) W-DBB and Time-Unrolled, serialized A-DBB allow the hardware to support activation density from $1/8$ to $8/8$, tuned on a per-layer basis,
implemented with a very cheap single MAC and 4:1 mux (DP1M4).
$M$ denotes index bitmask.
}
\label{fig:datapaths}
\end{figure*}

\Fig{fig:dbb_compress} gives a concrete DBB example, where the block size (BZ) is 8 and the maximum number of non-zero values (NNZ) per block is 4. In this particular case, the tensor blocking is performed along the channel dimension, which is a common strategy to avoid all the elements in any single channel falling into the same block.
The compression itself has two steps.
First, the non-zero elements are stored by removing the zeros.
Second, a simple bitmask $M$ is added to encode the presence of a non-zero element at each location in the expanded block.


We usually refer to a DBB block by the ratio $NNZ/BZ$. The example in \Fig{fig:dbb_compress} would be a 4/8 block. Note that any blocks that have less than NNZ non-zero elements will include one or more zeros in the compressed form.
\RED{For sparse DBB execution, a block cannot contain more than $NNZ$ non-zeros, achieved via pruning (Section ~\ref{sec:w-dbb} and~\ref{sec:a-dbb}). However, we also support a conventional dense mode for unpruned models.}



\subsection{Design Overview}
\label{sec:idea:ov}
\GAP

The rest of the paper discusses how to build an efficient accelerator to exploit weight and activation DBB sparsity, without significant accuracy drop. 
Weight sparsity is statically known, and thus the hardware design is relatively straightforward. Activation sparsity, however, is dynamic. We propose a dynamic pruning scheme co-designed with a novel time-unrolled architecture to exploit DBB sparsity in activations.

Building on top of DBB compression for both weight and activation, we describe how to integrate them into a complete DNN accelerator (\Sect{sec:npu}). In particular, we focus on the popular systolic architecture. We show that both weight and activation DBB support  can be easily integrated into a classic SA architecture by grouping PEs, leading to a tensor PE (TPE) design. Critically, the TPE design naturally exposes additional dimensions of data reuse that is unobtainable in the traditional SA design, enabling further efficiency gains.


We note that weight DBB sparsity has been used in a commercial A100 GPU from Nvidia~\cite{nv-a100-datasheet}. Our design differs in two ways. 
First, A100 supports only \RED{fixed 2/4 weight sparsity with up to 2$\times$ speedup, while our design comprehensively exploits both weight and variable activation DBB sparsity for up to 8$\times$ speedup using a novel time-unrolled architecture (Section~\ref{sec:adbb:arch}}). We show that activation DBB requires non-trivial extensions from previous weight DBB due to the dynamic nature of activation sparsity. 
Second, our design focuses on the systolic array architecture, which is well suited to low-power mobile/embedded systems. \RED{To fairly compare with A100, we implement an A100-featured systolic design, \mode{S2TA-W}, as a baseline in the evaluation.
Sec. \ref{sec:exp} and \ref{sec:eval:micro}}.

\section{Static Weight DBB Sparsity}
\label{sec:w-dbb}
\GAP

This section describes an architecture to apply DBB to weight sparsity (W-DBB). Note that weight DBB has been explored in \cite{kang-tcsvt19, SparseCore} and the proprietary A100 GPU from Nvidia with \RED{fixed 2/4 weight only W-DBB ~\cite{nv-a100-datasheet,A100-SP}}. Here, describing it allows us to explain the general principle of DBB sparsity, which we additionally apply to activations too.

The key advantage of weight DBB is that the load imbalance problem is greatly relaxed, as we have bounded the maximum number of non-zero elements per block and can provision hardware based on this.
We illustrate this by modifying the simple parallel (vector) architecture template given in Fig.~\ref{fig:datapaths}a, which show an 8-MAC dot product, which we refer to as DP8.
Building on top of DP8, Fig.~\ref{fig:datapaths}c shows a datapath for a $4/8$ W-DBB block, using only 4 hardware MACs instead of 8, yielding a 50\% reduction in MACs at the same throughput, and a 37.5\% reduction in weight operand bandwidth.
Since this design has only 4  MACs (DP4), with an 8-input mux (M8)\footnote{\RED{Omitting trivial optimization of mux widths for the sake of clarity.}}, we call this configuration DP4M8.

The key to the hardware for W-DBB is the 8:1 MUX in front of each MAC. The MUX, controlled by the positional bitmask($M$) from the weight DBB block, is used to steer the correct activation element into the MAC.
The overhead of this MUX is negligible, especially compared to the cost required to exploit unstructured sparsity (\Sect{sec:background}).
The DP4M8 (Fig.~\ref{fig:datapaths}c) also accommodates fall back to dense operation, which is essential to support models that have density greater than 50\%.


\section{Dynamic Activation DBB Sparsity}
\label{sec:a-dbb}
\GAP


Activation DBB compression (A-DBB) can be used together with W-DBB, allowing us to exploit the full potential of sparse data, without significant overheads.
However, A-DBB is more challenging than W-DBB because: (1) activation sparsity is unbounded, and (2) optimal activation DBB sparsity varies significantly across the layers of a network.
In this section, we describe the co-designed Dynamic Activation Pruning (DAP) and time-unrolled variable DBB architecture.


\subsection{Dynamic Activation Pruning (DAP)} 
\label{sec:a-dbb:dap}
\GAP

While weights can be compressed ahead of time (offline), activations are the result of runtime computation and therefore must be compressed online.
To do this, we propose Dynamic Activation Pruning (DAP),
which prunes and compresses the dense activation tensors into the DBB format. 
The random sparsity of a given activation tensor may have more non-zeros in a block than allowed by the DBB $NNZ$.
Therefore, DAP implements simple \textit{Top-NNZ} pruning to keep the block elements with the largest magnitude.
As with W-DBB (Fig.~\ref{fig:dbb_compress}), the activation tensor is first decomposed into 1\x 1\x $BZ$ blocks along the channel dimension.

DAP is a lossy scheme, and can degrade test accuracy on some models if $NNZ/BZ$ is small.
For example, 
MobileNetV1 shows a test accuracy drop from 71\% to 56.1\% when using $4/8$ DAP for all point-wise CNN layers.
We propose an extension to the conventional DNN training procedure to make up any accuracy loss from DAP (Sec~\ref{sec:eval:train}).

 

\begin{figure*}[!t]
\centering
\hspace*{-0.5cm}
\includegraphics[width=2.0\columnwidth]{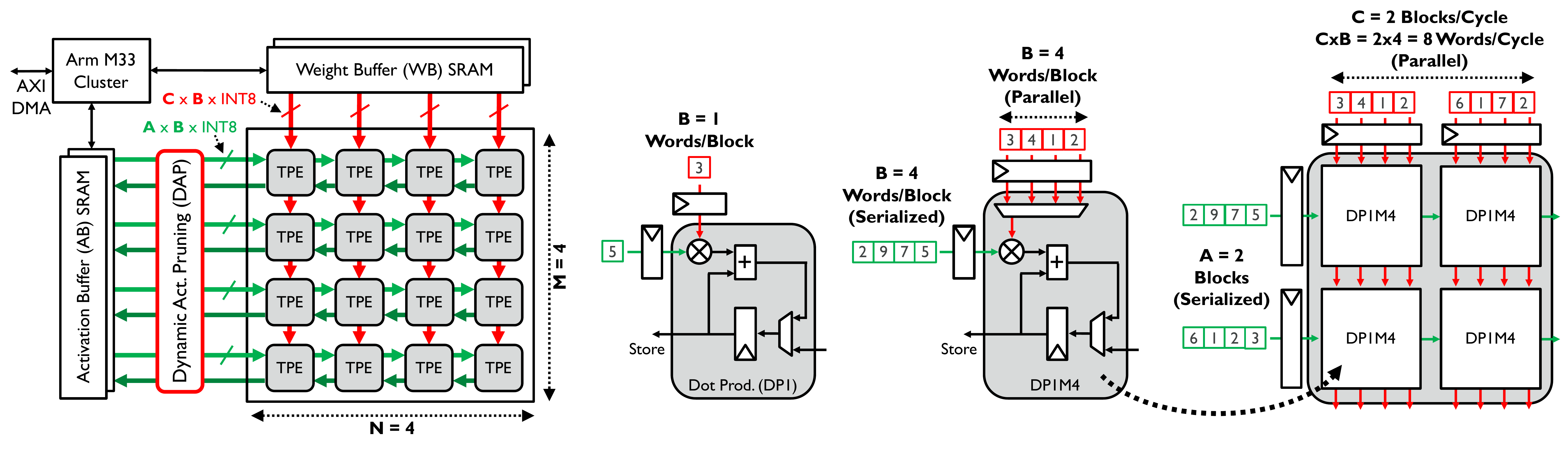}
\\
\hspace*{20pt}
(a) S2TA with M\x N TPE Array
\hspace*{37pt}
(b) 1\x 1\x 1 TPE
\hspace*{20pt}
(c) Time-Unrolled 2\x4\x 2 TPE with 4$\times$ DP1M4
\hspace*{10pt}
\vspace{8pt}
\caption{
Overview of (a) S2TA accelerator with TPE array (N$\times$M), dynamic activation pruning (DAP), local SRAM and Arm M33 MCUs.
\RED{Example TPE configurations are denoted A$\times$B$\times$C, where B is the $NNZ$ of the weight block.
A is the number of activation operand block inputs per cycle, and C is the number of weight block inputs.} 
TPE configs show for: (b) the classic scalar MAC PE (1\x 1\x 1), (c) A Time-Unrolled 2\x4\x 2 TPE implemented by connecting four DP1M4 units (\Fig{fig:datapaths}e).
}
\label{fig:system}
\end{figure*}

\subsection{Architectural Support for A/W-DBB}
\label{sec:adbb:arch}
\GAP

\paragraph{Basic Design} A simple implementation of joint DBB sparsity in a vector datapath is given in Fig.~\ref{fig:datapaths}d, where both weights and activations are compressed to reduce operand bandwidth.
As before, the only overheads are a single datapath multiplexer before each MAC, which this time is reduced to 4:1 (DP4M4).
The index bitmasks can be trivially compared to determine matching non-zero positions, and unused MACs can be clock gated to further reduce power.
Although this doesn't increase utilization of MACs, both weight and activations are now compressed in DBB format, thus the power to load both from SRAM is decreased dramatically.    

\paragraph{Challenges to Support Variable A-DBB} However, activation DBB introduces a significant challenge because the maximum sparsity varies significantly across layers, for example for ResNet50v1, 
the per-layer tuned activation DBB ranges from $8/8$ (dense) in early layers down to $2/8$ towards the end.
Therefore, while weight density typically hovers below 50\% and therefore $4/8$ W-DBB is a safe choice, forcing a fixed activation DBB sparsity (A-DBB) would be a huge compromise and result in diminished gains.
Therefore, 
it is necessary to support \textit{variable} A-DBB sparsity in the architecture.
Naively supporting variable DBB sparsity in hardware reduces utilization and leads to low energy and area efficiency.
For example, if we implement a $4/8$ DBB datapath and run a model with $2/8$ DBB, the utilization would drop by $\sim$50\%.



\paragraph{Time Unrolled Architecture for Variable A-DBB}
To support variable activation (A-DBB) sparsity efficiently, our key idea is to switch from unrolling the elements in the A-DBB blocks spatially (Fig.~\ref{fig:datapaths}d) to serializing them in time (Fig.~\ref{fig:datapaths}e).
This simply means that we process one element of the activation block per cycle using a single MAC, over multiple cycles.
For example, a layer with low density activations can use $1/8$ DBB, requiring just one cycle per block.
On the other hand, a layer with denser $5/8$ DBB activations requires five cycles.
In this way, we can now freely vary the activation block density per layer directly by changing the number of cycles per block, from $1/8$ to $8/8$.
Meanwhile, the datapath utilization and operand bandwidth remain constant. 
\vspace{-3pt}

\section{Structured Sparse Tensor Accelerator (S2TA)}
\label{sec:npu}
\GAP

This section describes how to integrate the architectural support for W-DBB and variable A-DBB into a DNN accelerator to exploit weight and activation sparsity. As a case-study, we focus on systolic array-based DNN accelerators. Due to their superior efficiency arising from local register-to-register operand reuse~\cite{kung_whysystolic}, the systolic array has been widely promoted for CNN accelerators~\cite{kung_2018,samajdar2018scale,scalesim-ispass20,smt-sa,nb-smt-micro20}, and even used in commercial products~\cite{tpu-isca}.
We call our DBB-exploiting systolic array \textit{S2TA}\RED{, with two main variants: (1) \mode{S2TA-W} as a comparison baseline targeting W-DBB alone, and (2) our optimal time-unrolled {\mode{S2TA-AW}} exploiting joint A/W-DBB}.

\subsection{Overall Architecture}
\label{sec:npu:arch}
\GAP


The key to supporting DBB sparsity in a systolic array is to replace the traditional \textit{scalar PE} with a \textit{Tensor PE (TPE)}. A scalar PE (Fig.~\ref{fig:system}b) accepts a single pair of operands per cycle and computes a single MAC.
A TPE (Fig.~\ref{fig:system}c), in contrast, accepts a pair of fixed size operand blocks per cycle.

TPE is a natural design choice to support DBB, since both weights and activations are naturally blocked in DBB. Each TPE essentially computes the MACs between an activation tensor and a weight tensor. 
\RED{One can implement the TPE using 
the DP4M8 MAC unit in \Fig{fig:datapaths}(c) to form \mode{S2TA-W}. For the time-unrolled \textbf{\mode{S2TA-AW}}}, the DP1M4 MAC unit in \Fig{fig:datapaths}(e) can be directly used as the TPE implementation.

A TPE can be configured in a variety of ways. 
For instance, \Fig{fig:system}(c) shows a TPE that accepts a weight tensor and an activation tensor \RED{in 4 cycles}, with size \RED{A$\times$B$^{'}$=2$\times$4 and B\x C=4\x2}, where B=4 is $NNZ$ of weight DBB blocks. 
The degenerate case of a 1\x1\x1 TPE is equivalent to the scalar PE from a traditional SA (Fig.~\ref{fig:system}b). Fig.~\ref{fig:system}c shows a 2$\times $4$\times$2 TPE to exploit joint A/W-DBB, which is implemented by connecting four time-unrolled scalar DP1M4 datapath in \Fig{fig:datapaths}(e) together in a pure \textit{outer-product} fashion. In contrast, a TPE implemented with DP4M8 datapath (\Fig{fig:datapaths}(\RED{c})) for exploiting W-DBB alone is reminiscent of a 4-way \textit{dot-product} \RED{with dense A\x B activation and sparse $NNZ$\x C weight tensor inputs to the TPE}.


At the array level, TPEs operate in the usual systolic fashion as shown in \Fig{fig:datapaths}a: each TPE receives a pair of tensor operands from adjacent neighbors and passes them on, except that the operands are tensors instead of scalar values.
\RED{Networks are mapped onto the array using simple matrix tiling, similar to the TPU~\cite{tpu-isca}, except we use output-stationary.}

\paragraph{Data Reuse} In addition to naturally supporting DBB sparsity, the TPE organization exposes two new dimensions of data reuse compared to the $1\times 1\times 1$ PE organization in classic systolic arrays. First, moving from a scalar MAC to a tensor-wise product in the TPE introduces \textit{intra-TPE accumulator reuse}, as we now achieve multiple MACs per accumulator update,
but lower the chance of ZVCG. Second, moving from a single scalar input to tensors
introduces \textit{intra-TPE operand reuse}, \RED{as each operand arriving at a TPE is used more than once. This amortizes the cost of moving operands across the array, amongst multiple MACs.}

These two new forms of data reuse result in much smaller on-chip buffer sizes, as the flip-flops required in the TPE are increasingly shared amongst a larger number of MAC units.
\RED{Table~\ref{tab:buffers} shows that 
S2TA-W with a 4\x4\x4\_4\x8 TPE array, and time-unrolled 8\x4\x4\_8\x8 S2TA-AW for $BZ$=8
have $\sim$7--1,886$\times$ less total buffers per MAC than previous architectures}. As a result, a larger TPE would also increase the energy efficiency, albeit at a marginally reduced clock frequency.
Note that the outer-product TPE is more efficient than the dot-product counterpart due to increased data reuse.

\subsection{Hardware Support for DAP}
\GAP

To support A-DBB, the hardware must implement DAP, which simply selects the $NNZ$ activation elements with the largest magnitude from the $BZ$-block (Section~\ref{sec:a-dbb:dap}).

\begin{figure}[t]
    \centering
    \includegraphics[width=0.5\textwidth]{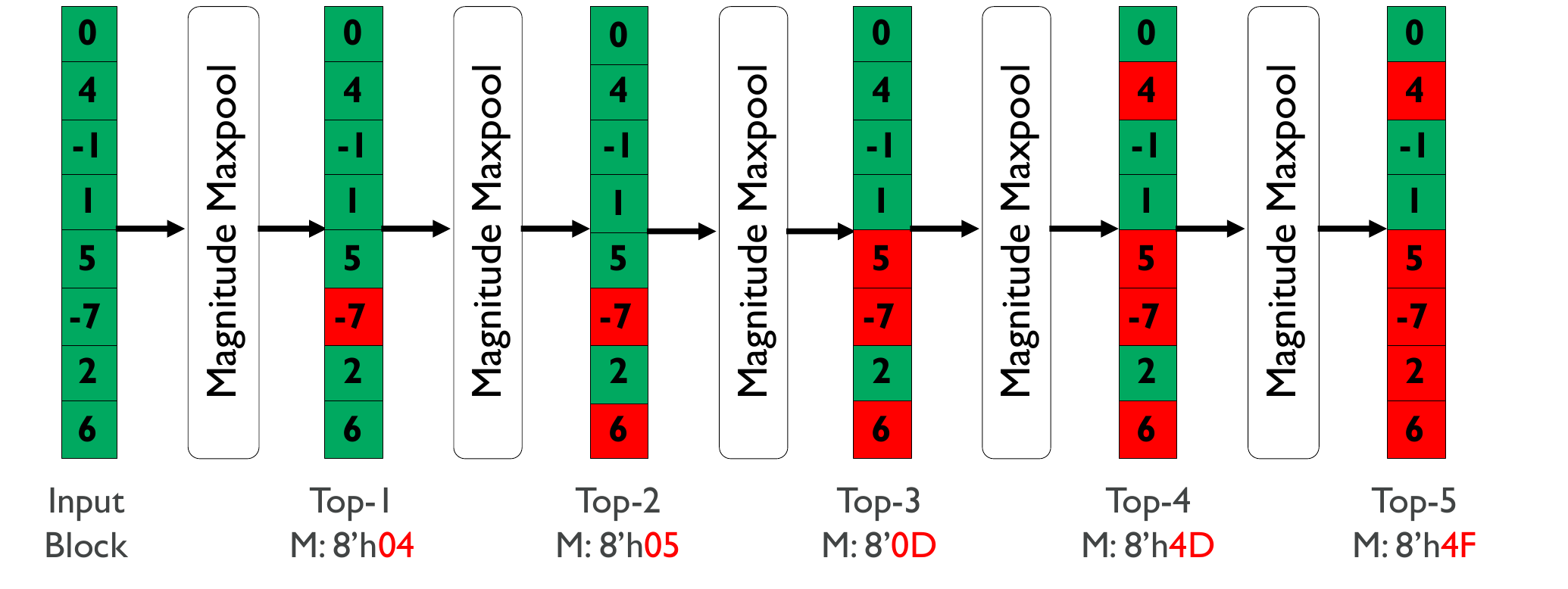}
    \caption{Hardware DAP array, consisting of cascaded magnitude (absolute value) maxpool stages, configured for an input block of $BZ$=$8$, with $NNZ$$\leq$$5$.
    The elements in red are discounted in consecutive maxpools. 
    For $4/8$ DBB, the output will be the elements $[4,5,-7,6]$ and the positional bit mask $M$=8$'$h4D.
    }
    \label{fig:dap5-hw}
\end{figure}

The challenge of providing hardware support for DAP is that the $NNZ$ in a A-DBB block is variable at run time. To support this, the DAP hardware cascades $NNZ$ number of maxpool stages. \Fig{fig:dap5-hw} shows the DAP hardware. Each maxpool selects the largest magnitude elements from the input block \RED{through binary comparison using $BZ$ - 1 comparators}, which in our current design is fixed to contain 8 elements as a block size of 8 is empirically found to provide a good balance between accuracy and efficiency (\Sect{sec:eval:train}).

We cap the maxpool stages at 5, since higher $NNZ$ would usually not lead to significant efficiency gains. Thus, the DAP hardware (Table \ref{tab:best_design}) supports a A-DBB sparsity ratio ranging from $1/8$ to $5/8$, \RED{bypassing any unused stages.}



\subsection{Other Design and Implementation Details}
\label{sec:npu:imp}
\GAP

\paragraph{On-chip SRAM} As is commonplace for accelerators, we heavily leverage local software managed SRAM~\cite{li-dac19} to provide a low-cost operand supply.
The 0.5MB weight buffer (WB) and the 2MB activation buffer (AB) are separate.
Both are double buffered to overlap computation in the TPE array and DMA data transfer.
The SRAM is grouped, rather than distributed, so we can use large high density SRAMs.

\paragraph{Local MCU with SIMD}
We implement non-GEMM operations such as activation functions, pooling, scaling, normalization and data type casting using Arm Cortex-M33~\cite{arm_m33} microcontrollers (MCUs), which have 32-bit SIMD instructions~\cite{arm_v8m}. M33 is very small (0.008mm$^2$~\cite{arm_m33}) and low power (3.9$\mu$W/MHz~\cite{arm_m33}) in 16 nm.
Control and data movement (DMA) tasks are also performed by the MCUs, e.g. loading the input image into AB.
We use a cluster of 4 MCUs each with a small 64KB control store SRAM, which is sufficient to ensure that the MCUs are never the performance bottleneck.

\section{Methodology}
\label{sec:exp}
\GAP

\paragraph{Automatic RTL Generation} The S2TA accelerator is highly modular and can be configured to make a calculated trade-off between area, performance, and power consumption. Instead of evaluating an arbitrary design point, we implement a parameterized Python RTL generator to explore the full design space, defined by five main parameters: the three TPE dimensions (A, B, C in \Sect{sec:npu:arch}) and the dimension of the entire SA (M, N); altogether denoted as 
A$\times$B$\times$C\_M$\times$N.
Each design can be further configured with 
any combination of W-DBB, A-DBB, ZVCG, and time-unrolling.

The RTL generator produces synthesizable Verilog RTL, along with a testbench suite. Each design is automatically validated in Synopsys VCS using the generated testbench, which executes inference on a given CNN model.
This generates accurate performance (throughput) metrics.
During the simulation, we also log value change dump (VCD) switching activity traces used for accurate annotated power simulation.

\begin{table}[t]
\caption{
Area and power breakdown of the S2TA-A/W design using a 16 nm process node,
8$\times$4$\times$\RED{4}\_\RED{8}$\times$8 TPE configuration for $BZ$=$8$, with 4 TOPS peak  throughput for 4/8 weight and dense activation.
}
\centering
\scriptsize
\begin{tabular}{l l l l l}
\toprule
\textbf{Component}                  &  \multicolumn{2}{l}{\textbf{Power, mW}}        & \multicolumn{2}{l}{\textbf{Area, mm$^2$}} \\
\midrule    
MAC Datapath and Buffers            &  317.7 & (58.7\%)             & 0.72  &    (19.1\%)  \\ 
Weight SRAM (512KB)                 &  69.4  & (12.8\%)             & 0.54  &    (14.3\%)   \\ 
Activation SRAM (2MB)               &  93.4  & (17.2\%)             & 2.16  &    (57.3\%)  \\
Cortex-M33 MCU~\cite{arm_m33} \x4   &  50.4  & (9.3\%)              & 0.30  &    (8.0\%)  \\ 
DAP Array                           &  10.4  & (2\%)                & 0.05  &    (1.3\%) \\
\midrule
Total                               &  541.3 & (100\%)              & 3.77  &    (100\%)  \\
\bottomrule
\end{tabular}
\label{tab:best_design}
\end{table}

\paragraph{Physical Design and Evaluation}
To evaluate area and power, each design goes through a complete EDA flow consisting of Synopsys and Cadence tools, with the TSMC PDKs and Arm multi-Vt cell libraries and single-ported SRAM compilers.
We used both TSMC 16nm FinFET and TSMC 65nm technology.
The clock frequency is constrained to 1GHz in 16nm (500MHz in 65nm) at the slow corner, with multiple process, voltage and temperature corners for setup and hold timing.
Power analysis was performed at the typical corner, using Synopsys PrimeTimePX, with the parasitic-annotated netlist and switching activity from VCD simulation traces.

\paragraph{S2TA-A/W Design Point} Based on typical mobile DNN accelerator specs, we set a hard constraint of 4 TOPS peak (dense) throughput and clock frequency of 1GHz in 16nm (500MHz in 65nm). We then sweep the design space to identify designs on the area-vs-power frontier, from which we identify the TPE array design with the lowest power,
which is the 8\x 4\x 4\_8\x 8 outer-product TPE implemented with the time-unrolled DP1M4 datapath (\Fig{fig:datapaths}(e)). This configuration with $4/8$ W-DBB and variable 
A-DBB is referred to as \mode{S2TA-AW}, and is used throughout the evaluation. Full area and power breakdown of \mode{S2TA-AW} is shown in Table~\ref{tab:best_design}.


\paragraph{Baselines} We compare against the following baselines:

\begin{itemize}
	\item \mode{SA-ZVCG}: classic SA design with zero-value clock gating (1\x 1\x 1\_32\x 64 TPE array in our notation).
	\item \mode{S2TA-W}: this is the variant of S2TA that exploits 4/8 W-DBB sparsity alone (dense activations), using a 4\x8\x4\_4\x8 TPE with 4-MAC dot-product datapaths (DP4M8 in \Fig{fig:datapaths}(c)).
	Comparing against \mode{S2TA-W} allows us to understand the gains from exploiting A-DBB and W-DBB jointly.
	This design also implements ZVCG to weakly exploit activation sparsity.
	\item \mode{SA-SMT}: our INT8 re-implementation of a recent systolic array design that exploits unstructured sparsity using operand staging FIFOs to distribute matching operands~\cite{smt-sa}, using the T2Q2 variant from the paper.
	\item \mode{SparTen} and \mode{Eyeriss-v2}: both are recent non-systolic array designs~\cite{sparten-micro19, eyeriss_v2} that exploit unstructured sparsity.
\end{itemize}

All systolic array designs have 4 TOPS peak (dense) throughput and otherwise identical configurations.
We use the same EDA flow to obtain performance, power, and area (PPA) metrics for a fair comparison. The PPA metrics for 
SparTen and Eyeriss-v2, are directly 
from the 
papers.

\section{Evaluation Results}
\label{sec:eval}
\subsection{Accuracy Results}
\label{sec:eval:train}
\GAP

As with virtually all forms of sparsity, DBB sparsity used in S2TA is lossy, and thus requires DNN fine-tuning to regain any accuracy loss. Here, we first describe the simple extensions to the conventional DNN training procedure to support DBB sparsity, followed by the accuracy results. We evaluate INT8 models since we focus on mobile inference, where INT8 is the most widely used for deployment~\cite{8bit-warden}.

\paragraph{Training for W-DBB} We apply magnitude based DBB-aware weight pruning, which is similar to random magnitude pruning~\cite{zhu2017prune}, but pruning independently within each DBB block. This typically runs for 20-50 epochs, progressively pruning small-magnitude weights within each DBB block, until the desired DBB sparsity constraint is met.

\Tbl{tab:dap-training} shows the results of W-DBB (with dense activation) fine-tuning of five popular CNNs with INT8 quantization: VGG-16, MobileNetV1, ResNet-50V1 on ImageNet, and LeNet-5 on MNIST.
We find that $4/8$ W-DBB density typically achieves $<$0.5\% accuracy loss, on both relatively big (ResNet-50V1) and compact 
(MobileNetV1) networks.

In general, a larger block size (BZ) relaxes accuracy loss, but increases the hardware cost to exploit the sparsity. Meanwhile, a larger NNZ per block increases accuracy while leaving less room for exploiting sparsity. This is evident as we increase the NNZ for ResNet-50V1 from 2 to 4 in \Tbl{tab:dap-training}. Overall, we find that $4/8$ DBB density level is a good compromise that achieves low accuracy loss for both compact and larger
models.
Compared to previous work, Kang~\cite{kang-tcsvt19} targets a fixed $2/8$ W-DBB, which is too aggressive to achieve good accuracy, while the Nvidia A100~\cite{nv-a100-datasheet} uses $2/4$, which is the same sparsity level as our $4/8$ choice, but less flexible.
In this paper, we apply BZ=8 throughout our description without losing generality.

\paragraph{Training for A-DBB} Dynamic Activation Pruning (DAP) is lossy and requires fine-tuning to minimize accuracy impact.
We incorporate DAP into DNN fine-tuning by adding DAP in front of convolution operations, mimicking how it is used at inference.
To back propagate through the DAP layer, we calculate the gradient of DAP with respect to the activation $a$ $\frac{\partial DAP(a)}{\partial a}$, which is a binary mask tensor with a value 1 for the \textit{Top-NNZ} elements and a value 0 for the pruned ones. 

As an example, MobileNetV1 shows a test accuracy drop from 71\% to 56.1\% when using $4/8$ A-DBB for all point-wise CNN layers before fine-tuning. A 30-epoch DAP-aware fine-tuning recovers the accuracy to 70.2\%.
We find that 50-100 epochs of fine-tuning are typically sufficient. 


\Tbl{tab:dap-training} also shows the accuracy of applying A-DBB alone and applying both forms of DBB sparsity jointly. Note that while W-DBB density is hypertuned on a per-model basis, the A-DBB density varies wildly from early layers to later layers and is therefore tuned per-layer (supported by S2TA-AW).
We observe that combined A/W-DBB has a 0.1\%--0.4\% accuracy loss compared to exploiting W-DBB alone.
\RED{Total additional training time is less than 48 hours in all cases.}

Finally, we demonstrate A/W-DBB pruning of Transformers, by training I-BERT~\cite{I-BERT} on the GLUE dataset \cite{wang-etal-2018-glue}.

\begin{table}[!t]
\caption{
Accuracies of the baseline (INT8) models and various DBB variants. The accuracy loss of exploiting both A-DBB and W-DBB is generally about 1\%. The accuracy loss of exploiting one form of sparsity is about 0.5\%. 
A-DBB density varies significantly across layers. We report the weighted average, which can be a non-integer ratio.
}
\vspace{2pt}
\centering
\scriptsize
\setlength\tabcolsep{2.2pt}
\begin{tabular}{>{\bfseries}l c c c c c c c }
\toprule
\textbf{Model}           & \textbf{Dataset}       & \textbf{Baseline$^3$} &  \multicolumn{3}{c}{----------- \textbf{DBB Pruning} -----------}      \\
                &               & \textbf{Acc. (\%)}  &  \textbf{A-DBB$^1$}  &  \textbf{W-DBB$^2$}  &   \textbf{Acc. (\%)}        \\
\midrule                            
LeNet-5        & MNIST         & 99.0             &  3/8     &   --        &  98.9      \\ 
LeNet-5        & MNIST         & 99.0              &  --  &   2/8       &  98.9      \\
LeNet-5        & MNIST         & 99.0              &  4/8     &   2/8      &  98.8      \\ 
\midrule                            
MobileNetV1         & ImageNet       & 70.1              &  3.8/8      &  --         &   69.4      \\
MobileNetV1         & ImageNet       & 70.1               &  --      &  4/8            &  69.8      \\
MobileNetV1 $^*$     & ImageNet      & 70.1              &  4.8/8     &  4/8            &   68.9    \\
\midrule   
AlexNet         & ImageNet      & 55.7             &  3.8/8    &  --        &  54.7 \\
AlexNet          & ImageNet      & 55.7             &  --      &  4/8        &  54.9 \\
AlexNet $^*$            & ImageNet      & 55.7             &  3.9/8    &  4/8       &  54.6 \\
\midrule   
VGG-16              & ImageNet       & 71.5              &    3.1/8       &  --   &   71.8      \\
VGG-16              & ImageNet       & 71.5              &    --          &  3/8   &   71.4      \\
VGG-16 $^*$         & ImageNet       & 71.5              &    3.1/8       &  3/8   &   {71.9}      \\
\midrule                            
ResNet-50V1         & ImageNet      & 75.0              &  -- &  4/8            & 74.5      \\
ResNet-50V1         & ImageNet      & 75.0              &  -- &  3/8            & 74.3      \\
ResNet-50V1         & ImageNet      & 75.0              &  -- &  2/8            & 73.1      \\
ResNet-50V1         & ImageNet      & 75.0              &  3.49/8 &  --   &     74.4  \\
ResNet-50V1 $^*$         & ImageNet      & 75.0              &  3.49/8    &  3/8   &     73.9  \\
ResNet-50V1              & ImageNet      & 75.0              &  3.49/8    &  4/8   &     74.1  \\
\midrule                            
{I-BERT (base)}  & {GLUE (QQP)}     &  {91.2}       &  {4/8 $^4$}     & \RED{--}     &    {91.2}             \\ 
{I-BERT (base)}  & {GLUE (QQP)}     &  {91.2}       &  {3/8 $^4$}     & \RED{--}     &    {91.0}             \\ 
{I-BERT (base)}  & {GLUE (QQP)}     &  {91.2}       &  {--}     &  4/8 $^4$   &    {91.1}             \\ 
{I-BERT (base)}  & {GLUE (QQP)}     &  {91.2}       &  {4/8 $^4$}     &  4/8 $^4$   &    {90.9}             \\ 
{I-BERT (base)}  & {GLUE (SST2)}     &  {94.7}       &  {4/8 $^4$}     & \RED{--}    &    {94.3}             \\ 
{I-BERT (base)}  & {GLUE (SST2)}     &  {94.7}       &  {4/8 $^4$}     & \RED{4/8 $^4$}    &    {93.5}             \\ 

\bottomrule
\end{tabular} 
\\
\vspace{2pt}
$^1$Tuned per-layer, average reported; and $-$ for dense.
$^2$Tuned per-model (excluding the 1st layer); and $-$ for dense.
$^3$8-bit dense models. 
{$^4$Fully-connected sub-layers (FC1, FC2) of Encoders.}  
$^*$ Used for whole model evaluation in \Sect{sec:eval:full}.
\label{tab:dap-training}
\end{table}

\begin{figure*}[t]
  \centering
  \hspace*{-0.15cm} 
  \subfloat[SA-ZVCG]
  {
  \includegraphics[width=.52\columnwidth]{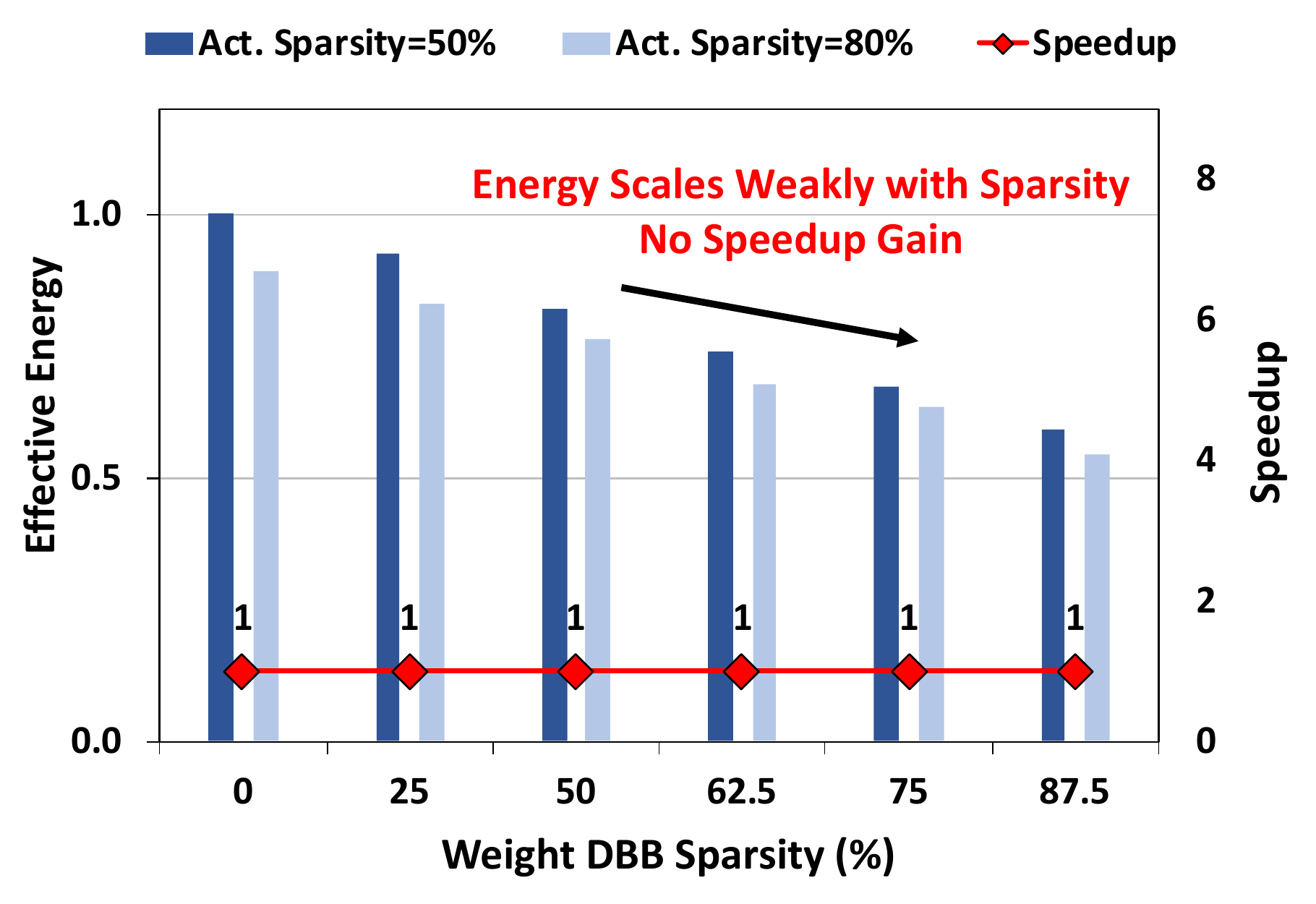}
  \label{fig:}
  }
  \hspace*{-0.3cm}
  \subfloat[{SA-SMT}]
  {
  \includegraphics[width=.52\columnwidth]{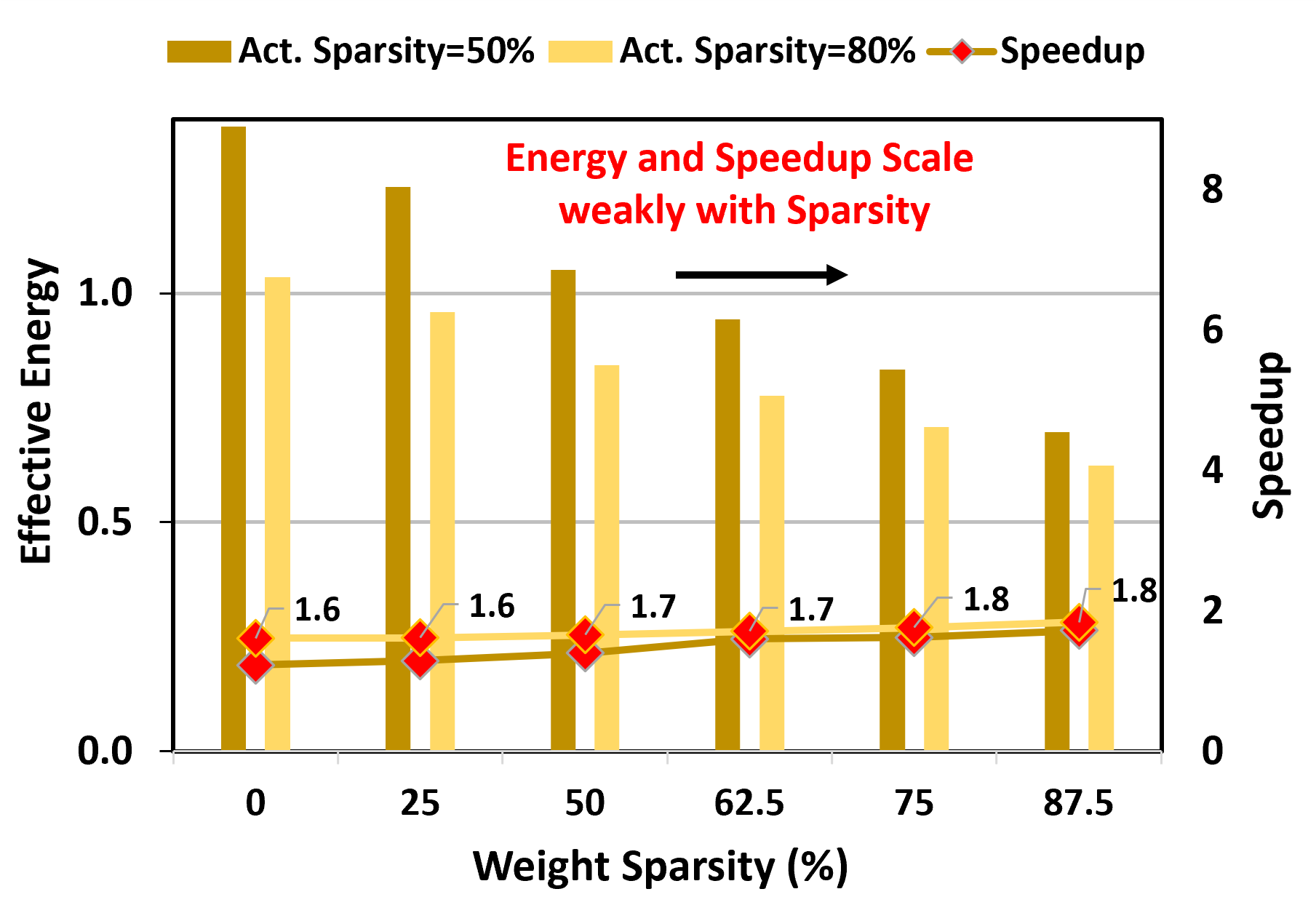}
  \label{fig:}
  }
  \hspace*{-0.3cm}
  \subfloat[S2TA-W]
  {
  \includegraphics[width=.50\columnwidth]{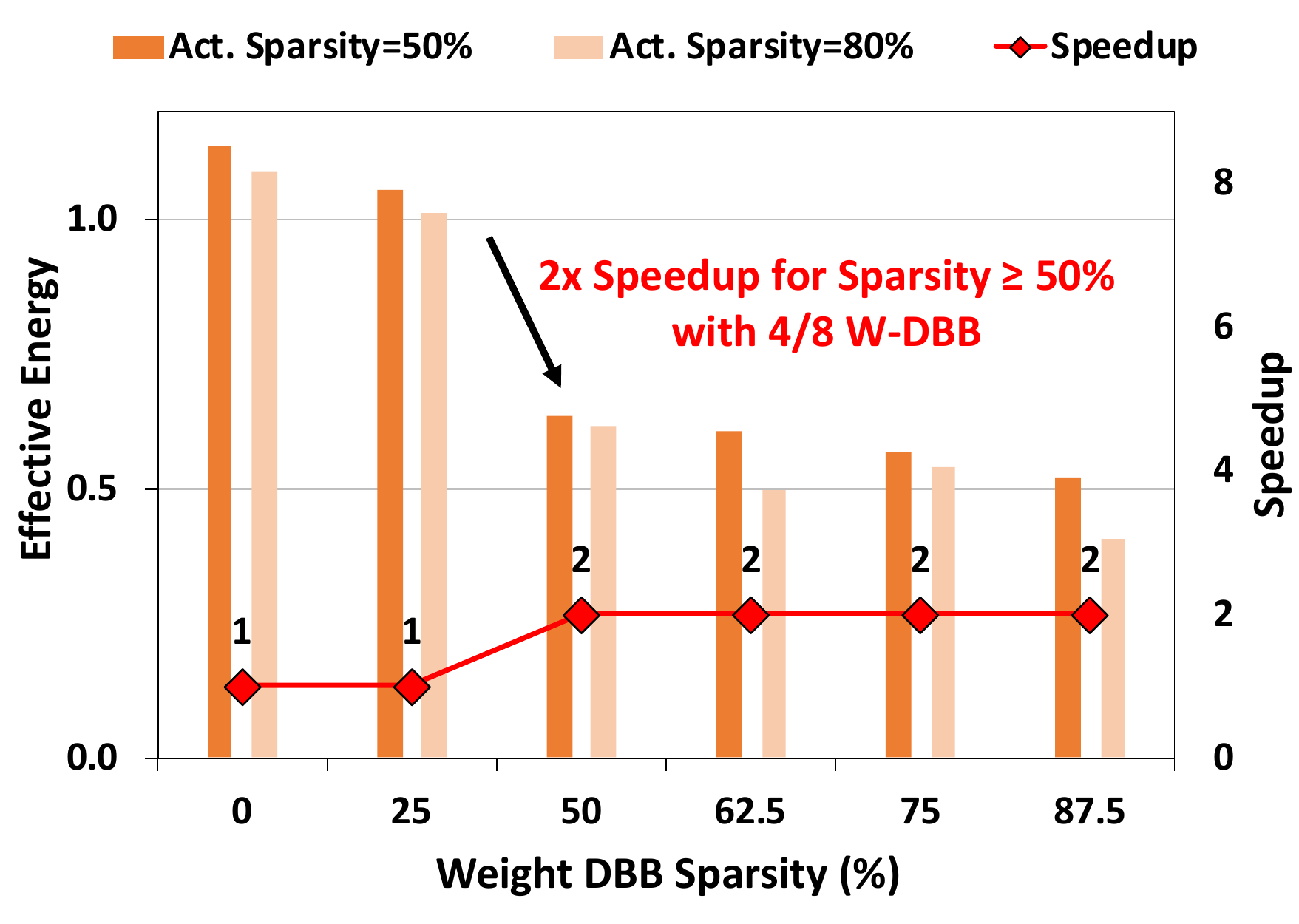}
  \label{fig:}
  }
  \hspace*{-0.3cm}
  \subfloat[S2TA-AW]
  {
  \includegraphics[width=.50\columnwidth]{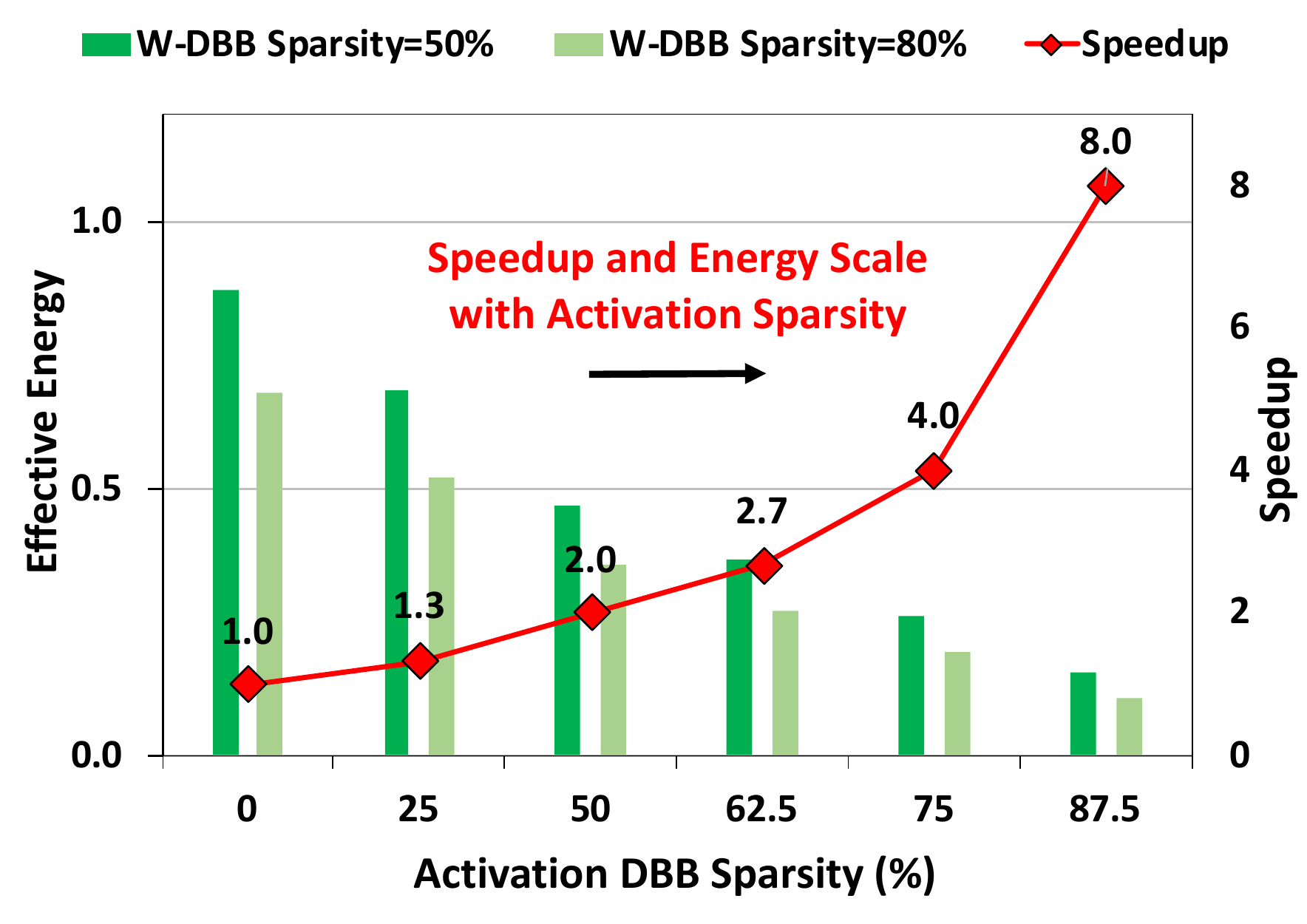}
  \label{fig:}
  }
  \caption{
Energy (normalized to \mode{SA-ZVCG} at 0\% activation and 50\% weight sparsity) and speedup with sparsity.
(a) \mode{SA-ZVCG} energy falls slowly with density, no speedup.
\RED{(b) \mode{SA-SMT} shows higher energy than \mode{SA-ZVCG}, up to 2$\times$ speedup.} 
(c) \mode{S2TA-W} exploiting W-DBB alone provides a fixed 2$\times$ speedup, and 1.2$\times$ energy reduction compared to \mode{SA-ZVCG}, at sparsity $\ge$50\%.
(d) \mode{S2TA-AW} with A/W-DBB exploits joint sparsity for up to 8$\times$ speedup and significant energy reduction.
}
\label{fig:sparsity}
\end{figure*}



\subsection{Microbenchmarking Results}
\label{sec:eval:micro} 
\GAP

\begin{figure}[!t]
    \centering
    \hspace*{-0.2cm}
    \includegraphics[width=1\columnwidth]{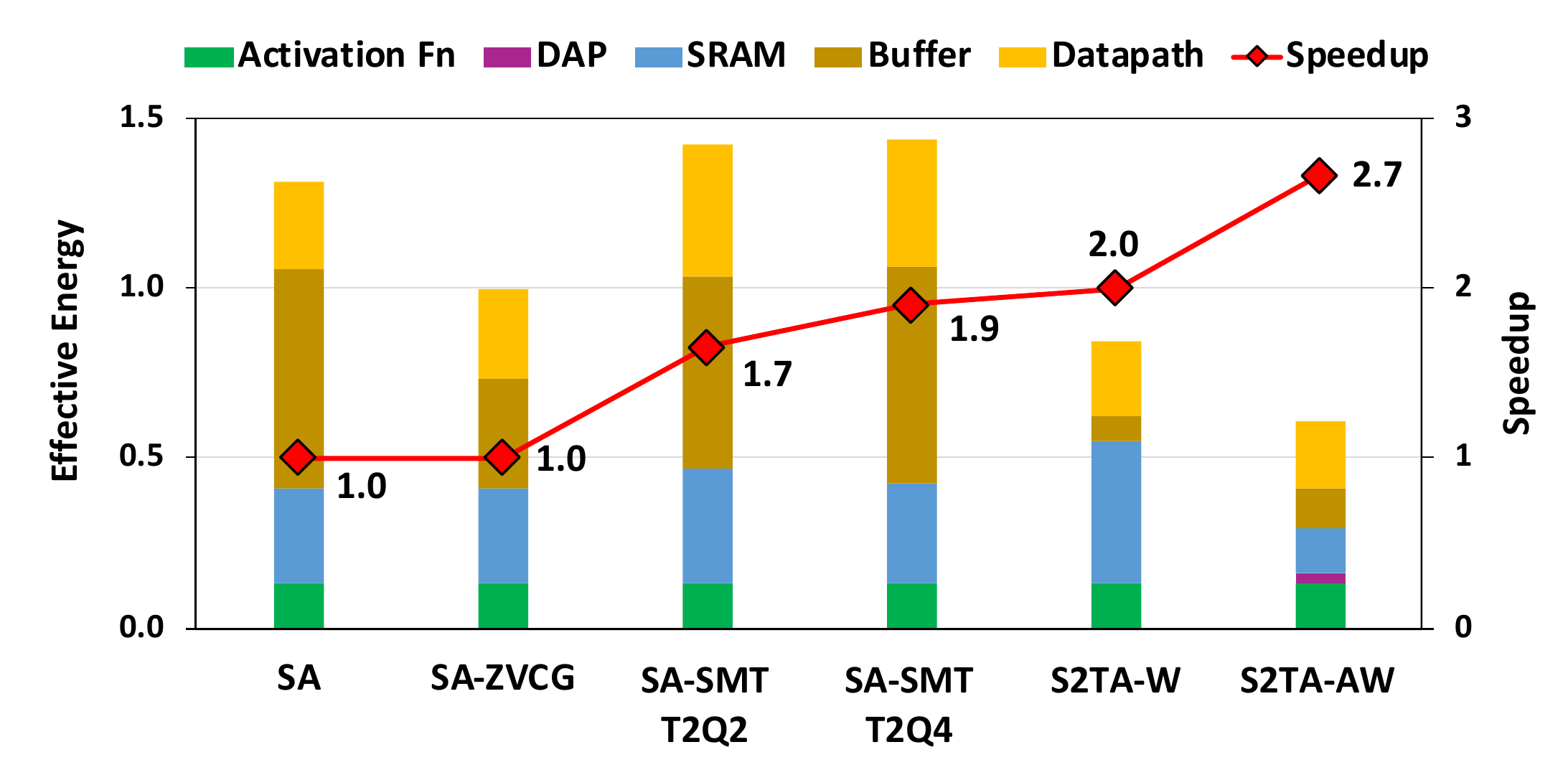}
    \caption{
    Energy breakdown and speedup of various SA variants for a typical convolution with 50\% ($4/8$-DBB) and 62.5\% ($3/8$-DBB) {INT8} weight and activation sparsity, respectively. The results are normalized to \mode{SA-ZVCG}. 
    }
    \label{fig:ppa-4-tops}
\end{figure}

We first use a set of synthetic (microbenchmark) DNNs with specific weight/activation sparsity to understand the performance, energy, and area of S2TA relative to the baselines.


\Fig{fig:sparsity}a show show the energy and performance of \mode{SA-ZVCG} vary with weight and activation density. The $x$-axis increases the weight sparsity from 0\% to 87.5\%, and the two bars refer to two different activation densities of 50\% and 20\%, respectively. The energy is normalized to energy per MAC operation. Naturally, the energy of \mode{SA-ZVCG} scales weakly as the weight and activation sparsity increases due to the clock gating, but there is no speedup regardless of the sparsity.
\RED{\Fig{fig:sparsity}b shows that \mode{SA-SMT} exploiting unstructured sparsity for both weight and activation consumes significant higher energy than the \mode{SA-ZVCG} in \Fig{fig:sparsity}a.}   

\paragraph{Exploiting W-DBB Alone} \Fig{fig:sparsity}\RED{c} shows the energy and performance of \mode{S2TA-W} normalized to \mode{SA-ZVCG}. \mode{S2TA-W} exploits a $4/8$ W-DBB sparsity and, thus, achieves a maximal 2$\times$ speedup step when weight sparsity is $\ge$50\%. 
The 2\x speedup also gives a corresponding energy reduction.
However, the energy reduction at 50\% weight sparsity then plateaus, only scaling weakly with additional DBB weight sparsity and lower switching activity.
Clearly, \mode{S2TA-W} cannot maximally exploit the abundant activation sparsity.

\paragraph{Exploiting A/W-DBB Jointly} 
Exploiting both forms of sparsity, \Fig{fig:sparsity}\RED{d} shows that \mode{S2TA-AW} achieves a 
significant energy reduction of up to 9.1$\times$ compared to \mode{SA-ZVCG}.

While the speedup from \mode{S2TA-W} is capped at 2$\times$ regardless of the activation sparsity, \mode{S2TA-AW} supports variable activation compression and so the speedup increases with activation sparsity from 1$\times$ at dense to 8$\times$ at 12.5\% activation density. 

\paragraph{Compared to Exploiting Unstructured Sparsity} 
\Fig{fig:ppa-4-tops} shows energy
and speedup 
on a typical convolution layer with 50\% weight and 62.5\% activation sparsity across all design variants.
\mode{SA-SMT}~\cite{smt-sa} is a SA-based accelerator exploiting unstructured sparsity using staging FIFOs with same INT8 operands.
The results are normalized to \mode{SA-ZVCG}. We break down the energy consumption into different components: the datapath of the PE array, the buffers in the PE array, SRAM, activation layers (M33), and the DAP logic, which is unique to \mode{S2TA-AW}.
We evaluate two variants of \mode{SA-SMT}: T2Q2 has an operand staging FIFO depth of two and T2Q4 has FIFO depth 4.
While both \mode{SA-SMT} variants are faster than \mode{SA-ZVCG}, they both also increase the effective energy consumption (43.0\% (T2Q2); 41.2\% (T2Q4)) versus the baseline \mode{SA-ZVCG}.
This is due to the energy overhead of the staging buffer for distributing matched operands, not necessary for \mode{S2TA-W} and \mode{S2TA-AW}.
This is also evident in the significantly lower buffer energy for \mode{S2TA-W} and \mode{S2TA-AW}.

\paragraph{A comparison between \mode{S2TA-W} and \mode{S2TA-AW}} \Fig{fig:ppa-4-tops} shows that the energy benefits of \mode{S2TA-AW} mainly come from a 3.1$\times$ reduction in the SRAM energy, as \mode{S2TA-AW} exploits activation sparsity using the time-unrolled outer-product TPE, whereas \mode{S2TA-W} loads dense activations from SRAM.

\begin{table*}[t]
\scriptsize
\caption{Comparison of \mode{S2TA-AW} and baselines, along with previously published sparse CNN accelerators in 16nm/65nm.
}
\vspace{1pt}
\centering
\scriptsize 
\setlength{\tabcolsep}{9pt} 
\begin{tabular}{>{\bfseries}l  l  c  c  c  c  c  c  c}
\toprule
    &                                                   & {SparTen}~\cite{sparten-micro19}   &  {Eyeriss v2}~\cite{eyeriss_v2} & {SA-ZVCG}$^1$ & {SA-SMT}$^{1}$~\cite{smt-sa}    
    & {S2TA-W}$^1$  & \textbf{\mode{S2TA-AW}}$^1$   \\
\midrule                    
\multicolumn{2}{l}{\textbf{Weight Sparsity}}                     & Random            & Random        &   ZVCG          & Random              & $\sfrac{4}{8}$ DBB, Dense     & $\sfrac{4}{8}$ DBB, Dense       \\
\multicolumn{2}{l}{\textbf{Activation Sparsity}}                 & Random            & Random        &   ZVCG          & Random              & Dense     & $\sfrac{(1-5)}{8}$ DBB, Dense       \\
\multicolumn{2}{l}{\textbf{SRAM Size (W/A)}}                     &     --            & 246KB         & 2MB / 0.5MB     & 2MB / 0.5MB         & 2MB / 0.5MB   & 2MB / 0.5MB                  \\
\multicolumn{2}{l}{\textbf{Hardware MACs}}                       & 32 (INT8)         & 384 (INT8)    & 2048 (INT8)       & 2048 (INT8)         & 2048 (INT8)     & 2048 (INT8)              \\
\midrule   
\textbf{Process Technology} &&\multicolumn{6}{c}{   
- - - - - - - - - - - - - - - - - - - - - - - - - - - - - - - -
16nm Implementations
- - - - - - - - - - - - - - - - - - - - - - - - - - - - - - - 
}  \\   
Clock Freq.                  & GHz                      &       --          & --            & 1.0             &   1.0             & 1.0          &  1.0                  \\
Area                         & mm$^2$                   &       --          & --            & 3.7             &   4.2             & 3.4          & 3.8                  \\
Peak Throughput$^2$          & TOPS                     &       --          & --            &  4              &   8               & 8    &   8 (16$^6$)              \\
Peak Energy Eff.$^2$         & TOPS/W                   &       --          & --            &  10.5 (12.8$^3$)&   \RED{8.01 (11.9$^3$)} & 12.4 (13.9$^3$)          &  \textbf{14.3 (26.5$^3$)}                \\
\cmidrule{1-1} 
\multirow{3}{*}{AlexNet}        
                             & $\times 10^3$ Inf./sec                &       --           & --            &  2.2 (3.0$^5$)          &    \RED{2.7 (4.0$^5$)} &   3.3 (5.0$^5$)        &   \textbf{3.7 (6.3$^5$)}       \\
                             & $\times 10^3$ Inf./J                  &       --       & --          &    4.81 (7.5$^5$)             &  \RED{4.48 (6.73$^5$)} & 7.2 (8.7$^5$)          & \textbf{9.9 (13.1$^5$)}   \\
                             & TOPS/W                          &   --     & --    &  7.2 (9.8$^5$)          &   6.7 (8.7$^5$)  &    9.8 (11$^5$)  &   \textbf{14 (19$^5$)}        \\
                              
\cmidrule{1-1}   
\multirow{3}{*}{MobileNet}        
                             & $\times 10^3$ Inf./sec                &       --           & --            &  2.7 (3.6$^5$)          &    \RED{3.5 (5.4$^5$)}     &    4.2 (7.3$^5$)     &   \textbf{5.2 (9.7$^5$)}         \\
                             & $\times 10^3$ Inf./J                  &       --          & --            &    7.9 (8.4$^5$)         &    \RED{7.5 (8.0$^5$)}     &    9.2 (9.9$^5$)           & \textbf{13.3 (14.9$^5$)}         \\
                             & TOPS/W                          &       --    & --      &  8.7 (9.2$^5$)     &    8.2 (9.0$^5$)     &    10 (11$^5$)       &   \textbf{14 (17$^5$)}        \\
\midrule   
\multicolumn{2}{l}{\textbf{Process Technology}}                  &     45nm           & 65nm          &  65nm           &        --         &     65nm        &   65nm                 \\
Clock Freq.                  & GHz                      &     0.8            & 0.2           &  0.5            &        --         &     0.5       &   0.5                  \\
Area                         & mm$^2$                   &     0.766\RED{$^6$}          & \RED{3.38$^{6,7}$}   &  21             &        --         &     --        &   24                   \\
Peak Throughput$^2$          & TOPS                     &      0.2           & 0.152         &   2             &         --        &     4         &   4 (8$^6$)                    \\
Peak Energy Eff.$^2$         & TOPS/W                   &      --            & --            &   0.78          &         --        &     0.87        &   \textbf{1.1}                 \\
\cmidrule{1-1}
\multirow{3}{*}{AlexNet}       
                             & $\times 10^3$ Inf./sec                 &       --           & 0.28 (0.34$^5$)       &  1.1 (1.5$^5$)      &     --      &    1.6 (2.5$^5$)     &   \textbf{1.8 (3.2$^5$)}        \\
                             & $\times 10^3$ Inf./J                   &       (0.52 $^5$)     & 0.66 (0.74$^5$)     &  0.44 (0.67$^5$)       &     --      &    0.55 (0.66$^5$)       &   \textbf{0.77 (1.02$^5$)}        \\
                             & \RED{TOPS/W}                           &       \RED{(0.68$^5$)}     & \RED{0.96 (1.1$^5$)}             &  \RED{0.67 (0.95$^5$)}        &     --      &    \RED{0.82 (0.95$^5$)}       &   \textbf{1.2 (1.4$^5$)}        \\
\cmidrule{1-1}
\multirow{3}{*}{MobileNet}       
                             & $\times 10^3$ Inf./sec                 &       --           & 0.13$^4$              &  1.6 (1.82$^5$)     &     --      &    1.82 (3.64$^5$)     &   \textbf{3.60 (4.85$^5$)}         \\
                             & $\times 10^3$ Inf./J                   &       --           & 0.22$^4$             &  0.68                 &     --      &    0.76                &   \textbf{1.04}       \\
                             & \RED{TOPS/W}                           &       \RED{--}     & \RED{0.24}           &  \RED{0.75}    &     --      &    \RED{0.83}       &   \textbf{1.1}        \\
\bottomrule
\end{tabular}
\\
\vspace{3pt}
$^1$Our results.
$^2$50\% sparse weights and activations.
$^3$75\% sparse weights and activations.
$^4$Scaled from MobileNet-v1-0.5-128 to 1.0-224. 
$^5$Conv only.
\RED{
$^6$Logic only.
$^7$2.7Mgates.
}
\label{tab:accelerator}
\end{table*}

\begin{figure}[t]
\centering
\includegraphics[width=0.46\textwidth]{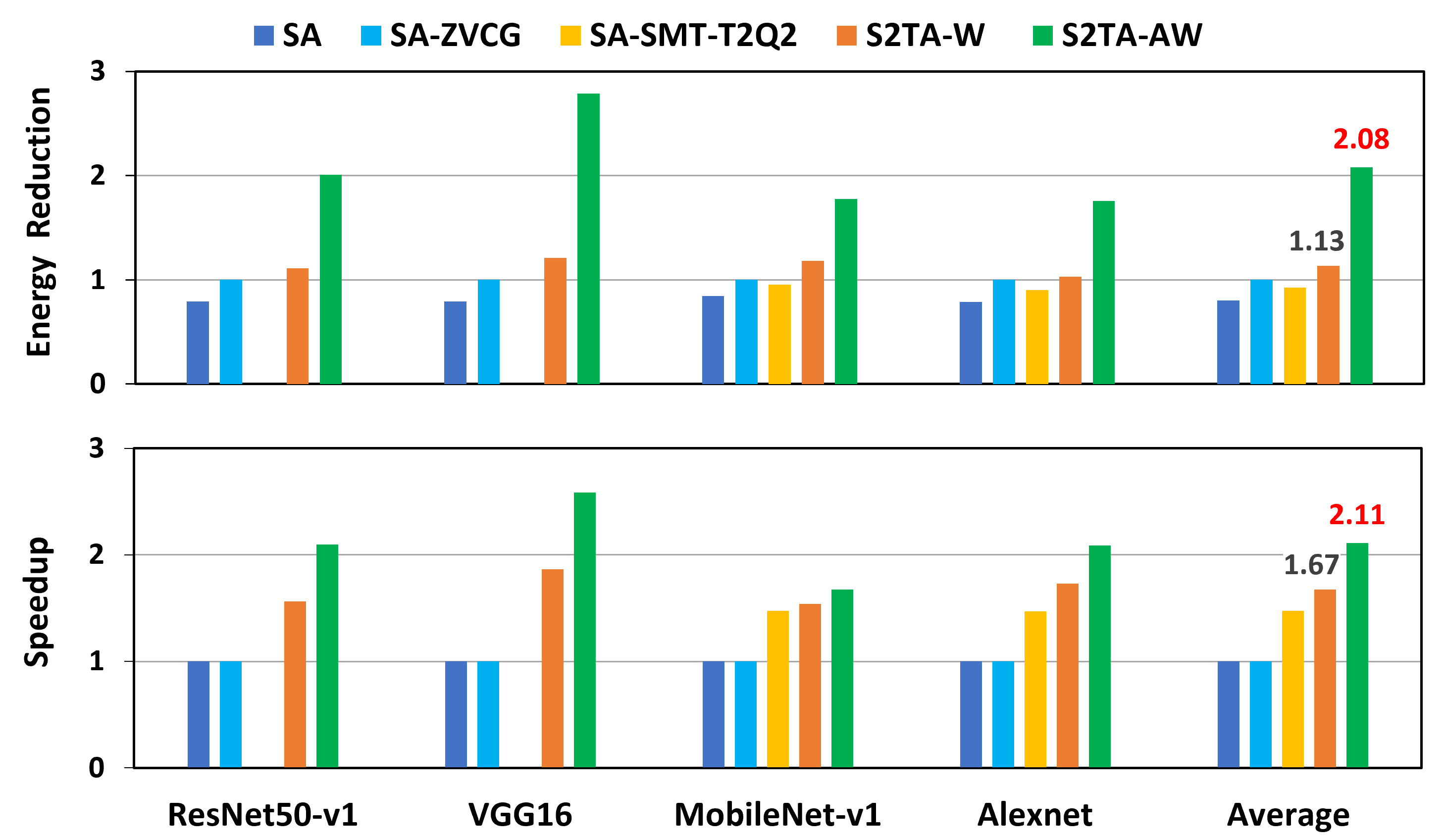}
\vspace{4pt}
\caption{
Energy reduction and speedup (normalized to \mode{SA-ZVCG}) comparison on ResNet50V1, VGG16,  MobileNetV1, and AlexNet (Convolution only) using the same 16nm technology.
The \mode{S2TA-AW} is 2.08$\times$, 1.84$\times$, \RED{2.24$\times$} energy efficient, and  2.11$\times$, 1.26$\times$, \RED{1.43$\times$} speedup than \mode{SA-ZVCG}, \mode{S2TA-W}, \RED{\mode{SA-SMT}} baselines, respectively.  
}
\label{fig:energy-latency-summary}
\end{figure}

\begin{figure}[!t]
\centering
\includegraphics[width=0.48\textwidth]{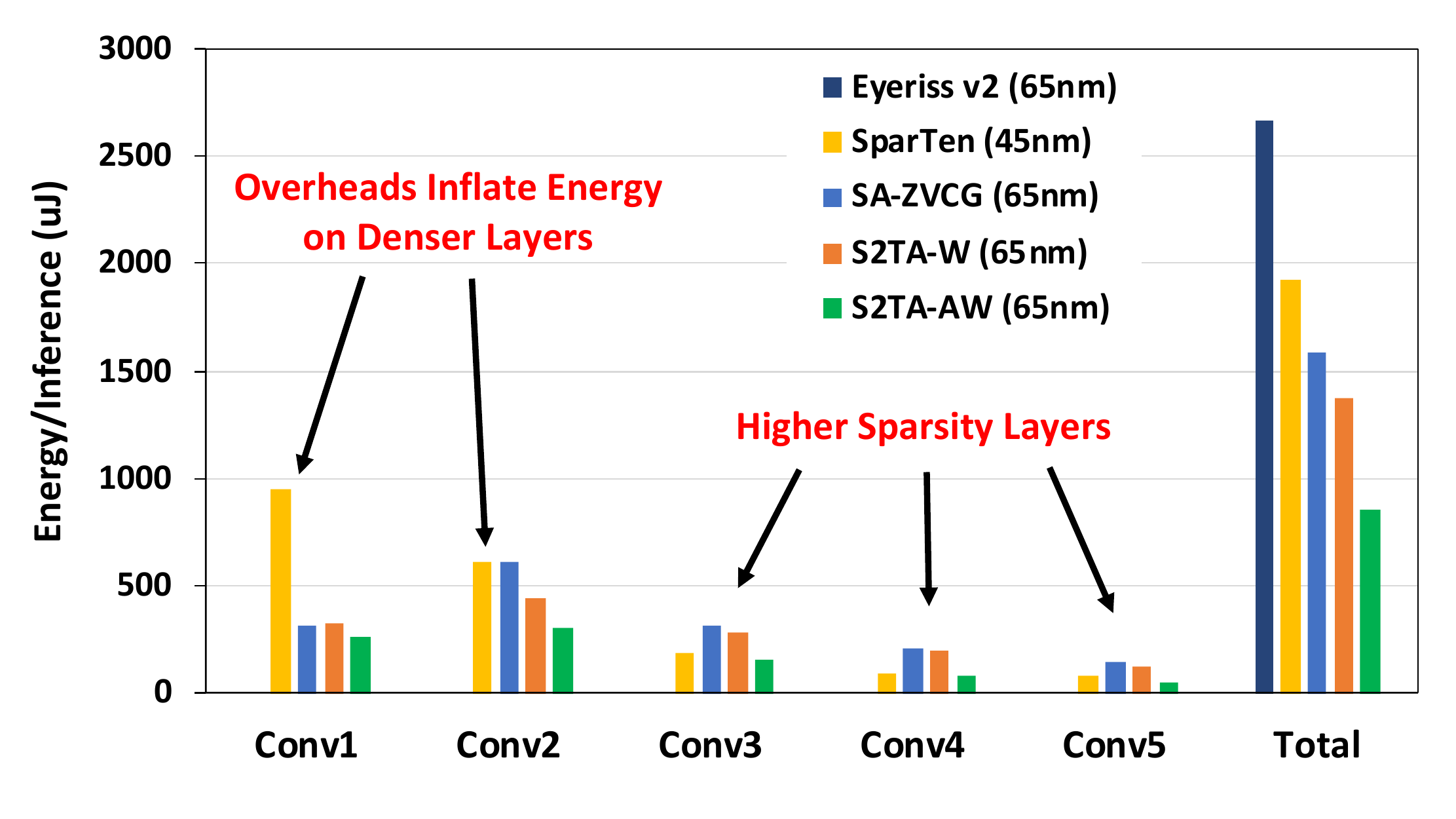}
\caption{
AlexNet per-layer energy for \mode{Eyeriss-v2}\cite{eyeriss_v2},  \mode{SparTen}~\cite{sparten-micro19}, \mode{SA-ZVCG}, \mode{S2TA-W}, and the optimal \mode{S2TA-AW}. 
\mode{S2TA-AW} in 65nm is about {2.2$\times$} more efficient than the 45nm \mode{SparTen} on AlexNet energy per inference.
SparTen has low energy only on very high sparsity layers (i.e. Conv3, 4 and 5). 
\mode{Eyeriss-v2} consumes 3.1$\times$ more total energy than \mode{S2TA-AW} in the same 65nm technology.
}
\label{fig:alexnet-vs-spartan}
\end{figure}

\subsection{Full Model Inference Results}
\label{sec:eval:full}
\GAP

We compare \mode{S2TA-AW} with \mode{SA}, \mode{SA-ZVCG}, \RED{\mode{SA-SMT}}, and \lword{\mode{S2TA-W}} 
across four models: VGG16, MobileNetv1, \lword{ResNet50v1}, and AlexNet. \Fig{fig:energy-latency-summary} gives energy reduction (top figure) and speedup (bottom figure) comparison, normalized to \mode{SA-ZVCG}. Compared to \mode{SA-ZVCG}, \mode{S2TA-AW} reduces energy by 1.76--2.79$\times$, with 1.67--2.58$\times$ speedup.

\paragraph{S2TA-W}
On average, \mode{S2TA-AW} consumes 1.84$\times$ lower energy and achieves {1.26}$\times$ speedup over \mode{S2TA-W} (Fig.~\ref{fig:energy-latency-summary}).





\paragraph{SA-SMT}
\mode{SA-SMT}~\cite{smt-sa} exploits unstructured sparsity in a systolic array.
\mode{SA-SMT} suffers from the overhead of distributing matching pairs, which can stall the data flow in the systolic array. \mode{SA-SMT} resolves it using expensive operand buffer FIFOs (\Sect{sec:background:oh1}).
We reimplemented \mode{SA-SMT}, which achieves 8.01 TOPS/W compared to 14.3 TOPS/W for \mode{S2TA-AW} at the same sparsity, as shown in \Tbl{tab:accelerator}.
This is due to the high energy cost of the FIFOs
(\RED{Fig.~\ref{fig:sparsity}b} and~\ref{fig:ppa-4-tops}).



\paragraph{Non-SA Accelerators} Next, we compare \mode{S2TA} with non-SA random-sparse accelerators, \mode{SparTen} and \mode{Eyeriss-v2}.

\mode{SparTen}~\cite{sparten-micro19} is a state-of-the-art CNN accelerator that exploits unstructured sparsity in weights and activations, with superior results to SCNN~\cite{scnn}.
For the comparison, we re-implemented \mode{S2TA-AW} in 65nm technology,
which is older than the 45nm used for \mode{SparTen} and, thus, does not unfairly favor \mode{S2TA-AW}.
\Fig{fig:alexnet-vs-spartan} compares the energy 
for \mode{S2TA-AW} and \mode{SparTen} on AlexNet, with 
\mode{SA-ZVCG} and \mode{S2TA-W}.
The total energy per inference on AlexNet is 2.2$\times$ lower on \mode{S2TA-AW} compared to \mode{SparTen}, even with a one node process disadvantage (65nm vs 45nm). The gain on MobileNet is even greater, as shown in \Tbl{tab:accelerator}. 
\mode{SparTen} performs best on layers with very high sparsity (e.g., Conv3-5), and less well on layers with dense or more moderate sparsity (e.g., Conv1,2).

A comparison with \mode{Eyeriss-v2}~\cite{eyeriss_v2} is given in \Tbl{tab:accelerator} and \Fig{fig:alexnet-vs-spartan}. \mode{S2TA-AW} shows $4.7\times$ and $1.4\times$ higher energy efficiency on MobileNet and AlexNet for convolution layers, respectively.
\Fig{fig:alexnet-vs-spartan} indicates \mode{Eyeriss-v2} is 3.1$\times$ less energy efficient than \mode{S2TA-AW} in the same 65 nm process technology.

As with any SA-based inference accelerators, fully-connected (FC) and depthwise (DW) layers are memory bound on S2TA, as batching is typically not used in inference.
However, we do prune FC/DW, included in the full model results in \Tbl{tab:accelerator}.


These results demonstrate that although the speedup gained by S2TA
is moderate, the overheads are low and the architecture is more energy efficient. In fact, we find that even the baseline \mode{SA-ZVCG} has lower energy than \mode{SparTen} on AlexNet and \mode{Eyeriss-v2} on MobileNet. While achieving high speedup from sparsity is important for certain scenarios, it is essential to also consider the overheads and start with an energy efficient architecture, which is especially important for mobile inference.

\subsection{Summary of Results}
\GAP

Finally, we summarize the key results, based on full-accelerator synthesis data with typical weight and activation sparsity.


\begin{enumerate}[leftmargin=*,itemsep=1pt]

\item DBB pruning of both weights and activations reduces the complexity of DNNs without significant accuracy loss (Table ~\ref{tab:dap-training}). 

\item \mode{SA-ZVCG} consumes 25\% less energy than a dense SA
by exploiting random sparsity.

\item
\mode{S2TA-W} exploiting W-DBB alone only marginally reduces energy (by 1.13$\times$) compared to \mode{SA-ZVCG};
exploiting both weight and activation sparsity is far superior.

\item
\mode{S2TA-AW} achieves an average 2.08$\times$ energy reduction and 2.11$\times$ speedup compared to 
\mode{SA-ZVCG} on full DNNs. 

\item 
Activation DBB sparsity varies widely across layers, thus the 
\mode{S2TA-AW} time-unrolled architecture supports variable 
activation DBB sparsity, with fixed weight sparsity. \footnote{\ S2TA time-unrolled architecture can also be implemented to support variable weight DBB sparsity and fixed activation DBB sparsity.} 

\item 
\mode{S2TA-AW} on typical CNN microbenchmarks with 50\% (75\%) weight and activation sparsity 
show 8 (16) TOPS, 14.3 (26.5) TOPS/W, and 2.16 (4.21) TOPS/mm$^2$, in 16nm, based on full-accelerator synthesis data in Table~\ref{table:designs-summary}.

\item
\mode{S2TA-AW} has 1.4 -- 4.7$\times$ lower energy than state-of-the-art accelerators exploiting unstructured sparsity, including \mode{SparTen}, \mode{Eyerissv2}, {and \mode{SA-SMT}} (Table~\ref{table:designs-summary}). 

\end{enumerate}
 
Table~\ref{table:designs-summary} summarizes key aspects of this work.

\begin{table*}[!htp]
\caption{
Summary of designs evaluated and previous works. S2TA-AW is a very low overhead fully-sparse architecture that achieves significant speedup and energy efficiency gains.
The optimal design is the Time-unrolled (Variable DBB) S2TA-AW architecture with up to 8$\times$ speedup.  
}
\vspace{1pt}
\centering
\scriptsize
\begin{tabular}{l c c c c c}
\toprule

\textbf{Architecture}  & \textbf{Weight Sparsity} & \textbf{Activation Sparsity} & \textbf{Hardware Overhead} & \textbf{ZVCG}  & \textbf{Variable DBB (Time-unrolling)} \\          
\midrule                                                           
\multicolumn{6}{c}{
\textit{
- - - - - - - - - 
Power Savings From Random Sparsity, No Speedup
- - - - - - - - - 
}} \\
SA~\cite{tpu-isca}                              &    \NO                       & \NO                        & --                               &  \NO                &  \NO                       \\
SA-ZVCG                                         &    \NO                       & \NO                        & --                               &  \YES               &  \NO                       \\
\midrule 
\multicolumn{6}{c}{
\textit{
- - - - - - - -
Speedup From Random Sparsity Incurs HW Overheads
- - - - - - - -
}} \\
SA-SMT~\cite{smt-sa}                            &    Random                    & Random                     & Gather                         &  \YES               &  \NO                       \\
SCNN~\cite{scnn}                                &    Random                   & Random                      & Scatter                        &  \NO                &  \NO                       \\
SparTen~\cite{sparten-micro19}                  &    Random                    & Random                     & Gather                           & \NO                & \NO                       \\
\midrule 
\multicolumn{6}{c}{
\textit{
- - - - - - - - - -
Speedup From Structured Sparsity: No Overheads
- - - - - - - - - -
}} \\
Kang~\cite{kang-tcsvt19}                        &    $\sfrac{2}{8}$ DBB        &   \NO                      & --                               &  \NO                &  \NO                       \\
STA~\cite{sta-cal2020}                          &    $\sfrac{4}{8}$ DBB        &   \NO                      & --                               &  \NO                &  \NO                       \\
A100~\cite{nv-a100-datasheet}                   &    $\sfrac{2}{4}$ DBB        &    \NO                     & --                               &  --$^1$             &  \NO                       \\
S2TA-W                                          &    $\sfrac{4}{8}$ DBB        &   \NO                      & --                               &  \NO                &  \NO                       \\
\textbf{S2TA-AW}                                &    $\sfrac{4}{8}$ DBB        &   {$\sfrac{(1-5)}{8}$ DBB}   & {--}                               &  \YES               &  \YES                    \\
\bottomrule
\end{tabular}
\\
\vspace{2pt}
$^1$ Unpublished proprietary design.
\label{table:designs-summary}
\vspace{5pt}
\end{table*}

\section{Related Work}
\label{sec:related}

\paragraph{Zero Value Clock Gating}
(ZVCG) (Sec.~\ref{sec:sparse_dnn}) saves power when 
having zero operands
~\cite{eyerissIsca,minerva,tpu-isca}.
We apply ZVCG to exploit excess sparsity that cannot be
exploited by DBB.

\paragraph{Indexed Unstructured Sparsity}
EIE~\cite{eie} implements a fine-grained
sparse CSR-encoded INT16 matrix-vector accelerator,
and ESE~\cite{ese_fpga17} extends this to LSTMs.
Doping~\cite{doping-mlsys2021} and MASR~\cite{masr_pact19} also exploit unstructured sparsity for LSTMs and RNNs, 
but uses a bitmask encoding.
A number of papers target unstructured sparse matrix multiplication for very sparse data, such as 
Outer Space~\cite{Outer-space}, which uses an outer product scheme, and SpArch~\cite{zhang2020sparch}, which further optimizes for locality.
Cnvlutin~\cite{albericio2016isca} skips compute for zero activations, without explicit indexes.
SCNN~\cite{scnn} implements a fully CSR-indexed sparse CNN accelerator using an outer product to exploit sparse weights and activations.
FixyNN~\cite{fixy2019sysml}
demonstrates a fixed-weight accelerator,
that can very efficiently exploit random sparsity.
SparTen~\cite{sparten-micro19} and Eyeriss v2~\cite{eyeriss_v2} both support fully-sparse inference.
We focus on DBB sparsity, but compare with SparTen, and Eyeriss v2
(Table~\ref{tab:accelerator}).

\paragraph{DBB Weight Sparsity} 
Kang~\cite{kang-tcsvt19} implements accelerator exploiting a fixed 2/8 W-DBB sparsity.
The design is based on a dot product microarchitecture with limited data reuse. 
Similar work\cite{SparseCore} also exploited W-DBB in the GPU context. 
The proprietary Nvidia A100 GPU implements fixed 2/4 W-DBB, 
which achieves 1.5$\times$ speedup and 3.12 TOPS/W (peak)\cite{SYSML20-Dally},
4$\times$ lower than the S2TA-W baseline at 12.4 TOPS/W (Table \ref{tab:accelerator}).

S2TA is the first architecture exploiting both W-DBB and A-DBB, and is the first to incorporate DBB into a systolic array with \RED{the novel time-unrolled technique} exploiting new dimensions of data reuse \RED{for up to 8$\times$ peak speedup}.

\paragraph{Sparsity in Systolic Arrays}
SAs (e.g. Google TPU~\cite{tpu-isca}) are efficient because they have high data reuse and local communication.
SMT-SA~\cite{smt-sa} is an SA that exploits unstructured FP32 sparsity using data staging FIFOs, which are energy inefficient for INT8 datapath, although acceptable for FP32.
Kung et al.~\cite{kung_2018} showed a preprocessing step of column combining of sparse weight matrices, before processing on a dense SA architecture.
Liu et al. \cite{sta-cal2020} exploited W-DBB sparsity for INT8 datapath in systolic architecture.   
NB-SMT~\cite{nb-smt-micro20} 
is a sparse SA with the ability to momentarily halve the MAC precision during to aid load balancing pipeline hazards.

\paragraph{Bit-wise Sparsity}
Even sign extension and zero bits within a word can be considered for optimization:
Pragmatic ~\cite{pragmatic_micro17} 
implements weight bit-sparsity, Tactical ~\cite{bittactical_asplos19}
implements activation bit-sparsity, and Laconic ~\cite{laconic_isca19} 
implements weight and activation bit-sparsity.
While orthogonal to our work, bit-sparsity is an interesting avenue for future DBB research.

\section{Conclusion}
\label{sec:conclusion}
\GAP

Density bound block (DBB) 
exploits structured sparsity, without the overheads of random sparsity schemes.
We describe an architecture to aggressively exploit DBB sparsity on both weights and activations.
For weight DBB, we 
prune during training to 
meet the DBB structured sparsity constraint.
However, we cannot do this for activations, which are not static, but rather computed at runtime.
Therefore,
we introduce 
Dynamic Activation Pruning (DAP),
a co-design solution for activation DBB, 
which implements activation DBB in hardware during runtime to force the required DBB sparsity.
DAP 
is lossy, and therefore must also be incorporated during 
training
to prevent accuracy loss at inference time.


The proposed novel time-unrolled DBB sparsity architecture, S2TA-AW, implements joint weight and activation sparsity in an efficient systolic architecture. The design significantly outperforms other 
strong baselines,
including the variants S2TA-W \RED{and SA-SMT}, which exploit weight sparsity alone, \RED{and fully unstructured sparsity, respectively}. 
On the pruned INT8 benchmark models AlexNet, MobileNetv1, VGG16, ResNet50v1, a 16nm S2TA-AW with joint weight and activation DBB sparsity demonstrates 2.08$\times$ and 1.84$\times$ energy reduction, on average, compared to the baseline SA-ZVCG and S2TA-W which exploits weight DBB sparsity alone, respectively.
Finally, S2TA-AW has about 2$\times$ and 3$\times$ the energy efficiency of the state-of-the-art unstructured sparse accelerator SparTen and Eyersiss-v2 on AlexNet, respectively.
\newpage

\bibliographystyle{IEEEtranS}
\bibliography{refs_vista}

\end{document}